\documentclass[12pt]{article}
\usepackage{epsfig}

\usepackage{mathrsfs}
\usepackage[T1]{fontenc}
\usepackage{mathpazo}
\usepackage{setspace}
\usepackage{amsfonts}
\usepackage{amssymb}
\usepackage{amsmath}
\usepackage{epsfig}
\usepackage{latexsym}
\usepackage{color}
\usepackage{graphicx}
\usepackage{nicefrac}
\usepackage[latin1]{inputenc}
\usepackage{pstricks}
\usepackage{slashed}

\def\hybrid{\topmargin -20pt    \oddsidemargin 0pt
        \headheight 0pt \headsep 0pt
        \textwidth 6.25in       
        \textheight 9.5in       
        \marginparwidth .875in
        \parskip 5pt plus 1pt   \jot = 1.5ex}

\hybrid

\catcode`\@=11

\def\marginnote#1{}
\newcount\hour
\newcount\minute
\newtoks\amorpm
\hour=\time\divide\hour by60 \minute=\time{\multiply\hour by60
\global\advance\minute by-\hour}
\edef\standardtime{{\ifnum\hour<12 \global\amorpm={am}%
        \else\global\amorpm={pm}\advance\hour by-12 \fi
        \ifnum\hour=0 \hour=12 \fi
        \number\hour:\ifnum\minute<10 0\fi\number\minute\the\amorpm}}
\edef\militarytime{\number\hour:\ifnum\minute<10
0\fi\number\minute}

\def\draftlabel#1{{\@bsphack\if@filesw {\let\thepage\relax
   \xdef\@gtempa{\write\@auxout{\string
      \newlabel{#1}{{\@currentlabel}{\thepage}}}}}\@gtempa
   \if@nobreak \ifvmode\nobreak\fi\fi\fi\@esphack}
        \gdef\@eqnlabel{#1}}
\def\@eqnlabel{}
\def\@vacuum{}
\def\draftmarginnote#1{\marginpar{\raggedright\scriptsize\tt#1}}

\def\draft{\oddsidemargin -.5truein
        \def\@oddfoot{\sl preliminary draft \hfil
        \rm\thepage\hfil\sl\today\quad\militarytime}
        \let\@evenfoot\@oddfoot \overfullrule 3pt
        \let\label=\draftlabel
        \let\marginnote=\draftmarginnote
   \def\@eqnnum{(\theequation)\rlap{\kern\marginparsep\tt\@eqnlabel}%
\global\let\@eqnlabel\@vacuum}  }

\def\preprint{\twocolumn\sloppy\flushbottom\parindent 2em
        \leftmargini 2em\leftmarginv .5em\leftmarginvi .5em
        \oddsidemargin -.5in    \evensidemargin -.5in
        \columnsep .4in \footheight 0pt
        \textwidth 10.in        \topmargin  -.4in
        \headheight 12pt \topskip .4in
        \textheight 6.9in \footskip 0pt
        \def\@oddhead{\thepage\hfil\addtocounter{page}{1}\thepage}
        \let\@evenhead\@oddhead \def\@oddfoot{} \def\@evenfoot{} }

\def\numberbysection{\@addtoreset{equation}{section}
        \def\theequation{\thesection.\arabic{equation}}}

\def\underline#1{\relax\ifmmode\@@underline#1\else
        $\@@underline{\hbox{#1}}$\relax\fi}

\def\titlepage{\@restonecolfalse\if@twocolumn\@restonecoltrue\onecolumn
     \else \newpage \fi \thispagestyle{empty}\c@page\z@
        \def\thefootnote{\fnsymbol{footnote}} }

\def\endtitlepage{\if@restonecol\twocolumn \else \newpage \fi
        \def\thefootnote{\arabic{footnote}}
        \setcounter{footnote}{0}}  

\catcode`@=12 \relax

\def\figcap{\section*{Figure Captions\markboth
        {FIGURECAPTIONS}{FIGURECAPTIONS}}\list
        {Figure \arabic{enumi}:\hfill}{\settowidth\labelwidth{Figure
999:}
        \leftmargin\labelwidth
        \advance\leftmargin\labelsep\usecounter{enumi}}}
 \relax
\def\tablecap{\section*{Table Captions\markboth
        {TABLECAPTIONS}{TABLECAPTIONS}}\list
        {Table \arabic{enumi}:\hfill}{\settowidth\labelwidth{Table
999:}
        \leftmargin\labelwidth
        \advance\leftmargin\labelsep\usecounter{enumi}}}
 \relax
\def\reflist{\section*{References\markboth
        {REFLIST}{REFLIST}}\list
        {[\arabic{enumi}]\hfill}{\settowidth\labelwidth{[999]}
        \leftmargin\labelwidth
        \advance\leftmargin\labelsep\usecounter{enumi}}}
 \relax
\makeatletter
\newcounter{pubctr}
\def\publist{\@ifnextchar[{\@publist}{\@@publist}}
\def\@publist[#1]{\list
        {[\arabic{pubctr}]\hfill}{\settowidth\labelwidth{[999]}
        \leftmargin\labelwidth
        \advance\leftmargin\labelsep
        \@nmbrlisttrue\def\@listctr{pubctr}
        \setcounter{pubctr}{#1}\addtocounter{pubctr}{-1}}}
\def\@@publist{\list
        {[\arabic{pubctr}]\hfill}{\settowidth\labelwidth{[999]}
        \leftmargin\labelwidth
        \advance\leftmargin\labelsep
        \@nmbrlisttrue\def\@listctr{pubctr}}}
 \relax
\makeatother
\newskip\humongous \humongous=0pt plus 1000pt minus 1000pt

\newif\ifdtup

\relax

\def\be{\begin{equation}}
\def\ee{\end{equation}}
\def\ba{\begin{eqnarray}}
\def\ea{\end{eqnarray}}

\newcommand{\tri}{\hspace{-3.5pt}\vartriangle\hspace{-3.5pt}}

\renewcommand{\theequation}{\thesection.\arabic{equation}}

\newcommand{\eqn}[1]{(\ref{#1})}

\author{
  \begin{minipage}{.97\linewidth}
    \vspace{0cm}
    \begin{center}
      \begin{small}
        \textbf{Ioannis Bakas}\footnote{bakas@mail.ntua.gr} ${\ }^1$ and
               \textbf{Dieter L\"ust}\footnote{dieter.luest@lmu.de} ${\ }^{2,3}$
        \end{small}
    \end{center}
    \vspace{0.5cm}
    \hspace{1.5cm}\begin{minipage}{.8\linewidth}
     {\it \begin{footnotesize}
    \begin{itemize}
        \item[${}^1$] Department of Physics, School of Applied Mathematics and Physical Sciences \\
        National Technical University, 15780 Athens, Greece
      \item[${}^2$] Max-Planck-Institut f\"ur Physik\\
       F\"ohringer Ring 6, 80805 M\"unchen, Germany
                    \item[${}^3$] Arnold-Sommerfeld-Center f\"ur Theoretische Physik\\
        Department f\"ur Physik, Ludwig-Maximilians-Universit\"at M\"unchen\\
        Theresienstra\ss e 37, 80333 M\"unchen, Germany
       \end{itemize}
     \end{footnotesize}}
    \end{minipage}
    \vspace{0.5cm}
  \end{minipage}
}

\title{\vspace{0.6cm}
 \boldmath \begin{LARGE}
    \textbf{3-Cocycles, Non-Associative Star-Products and\\
    the Magnetic Paradigm of $R$-Flux String Vacua}
  \end{LARGE} \unboldmath
}

\begin{document}

\renewcommand{\thepage}{\arabic{page}}
\setcounter{page}{1}


\begin{titlepage}
  \maketitle
  \thispagestyle{empty}

  \vspace{-13.9cm}
  \begin{flushright}
    LMU-ASC 64/13\\
    MPP-2013-255
  \end{flushright}

  \vspace{11cm}

  \begin{center}
    \textsc{Abstract}\\
  \end{center}
 We consider the geometric and non-geometric faces of closed string vacua arising by T-duality from principal torus bundles
 with constant $H$-flux and pay attention to their double phase space
 description encompassing all toroidal coordinates, momenta and their dual on equal footing.
 We construct a star-product algebra on functions in phase space that is manifestly duality invariant and substitutes for
 canonical quantization. The 3-cocycles of the
 Abelian group of translations in double phase space are seen to account for non-associativity of the star-product.
 We also provide alternative cohomological descriptions of non-associativity and draw analogies with the quantization of
 point-particles in the field of a Dirac monopole or other distributions of magnetic charge. The magnetic field analogue of the
 $R$-flux string model is provided by a constant uniform distribution of magnetic charge in space and non-associativity
 manifests as breaking of angular symmetry. The Poincar\'e vector comes to rescue angular symmetry as well as associativity
 and also allow for quantization in terms of operators and Hilbert space only in the case of charged particles
 moving in the field of a single magnetic monopole.

\end{titlepage}

\tableofcontents
\newpage


\section{Introduction}
\setcounter{equation}{0}

Flux backgrounds received considerable attention in recent years, while developing superstring
theory and its viable phenomenological applications to model building for elementary
particle physics (for reviews see \cite{Grana:2005jc,Blumenhagen:2006ci}).
Typically, the fluxes act as stabilizing factors leading to non-trivial
fixed points of the renormalization group equations, which would not have existed otherwise. Then,
T-duality transformations in the presence of fluxes, such as Neveu-Schwarz fluxes associated to
a closed 3-form $H$, which will be considered throughout this paper, were used to exhibit that not
only the geometry but also the topology of space in which strings propagate are not perceived
as in ordinary point-particle theories. In particular, for $S^1$ fibrations over a base manifold $M$,
i.e., $S^1 \rightarrow X \rightarrow M$,
T-duality interchanges the fibrewise integral of the $H$-flux with the first Chern class of the
bundle $X$. Another notable example of the change of topology led by fluxes is string theory on
the lens space $S^3/\mathbb{Z}_n$ with $m$ units of $H$-flux that is T-dual to string theory on
$S^3/\mathbb{Z}_m$ with $n$ units of $H$-flux. Untwisting $S^3$ to $S^2 \times S^1$ by one unit
of flux is a simple instance of this equivalence.

Dimensional reduction and $T$-duality of flux
backgrounds $X$ that are $T^n$ torus fibrations over a base manifold $M$ that may also depend on $n$,
$T^n \rightarrow X \rightarrow M$, encompass many more possibilities and can lead to novelties
that we are just beginning to comprehend and study their implications to theory and phenomenology.
In particular, for $n=2$ (hereafter called non-geometric $Q$-flux background), T-dualities can induce fuzziness,
apart from topology change, leading to non-commutative tori in closed string compactifications
\cite{Lust:2010iy,Condeescu:2012sp,Andriot:2012vb}, whereas for $n \geq 3$ (the non-geometric $R$-flux background)
the situation becomes even more interesting as non-geometric closed string backgrounds attached to non-associative tori
come into play \cite{Blumenhagen:2010hj,Blumenhagen:2011ph}. Similar results based on somewhat less explicit
methods appeared in the literature before \cite{bouwknegt1, bouwknegt2, bouwknegt3, bouwknegt4, mathai},
while studying the action of T-duality on torus fibrations with fluxes.

The emergence of novel mathematical structures from flux backgrounds calls for better understanding
of the issues involved, using different view-points, some of which will be presented in this paper.
Without loss of great generality, and in order to avoid unnecessary technicalities, it suffices
to formulate the problem by taking the base manifold $M$ as a point and consider the torus $T^3$
in the presence of constant $H$-flux. Then, one can think of $T^3$ as circle fibration over $T^2$
and apply the usual duality rules to obtain a twisted torus with no flux. Alternatively, one can think
of $T^3$ as a $T^2$ fibration over $S^1$ and perform one more duality. The resulting closed string
background (the $Q$-flux model) is only locally geometric, but not globally, since the
transition functions between two coordinate patches are prescribed in terms of T-duality transformations
and not in terms of diffeomorphisms. Then, because of non-trivial monodromies characterizing the $T^2$
fibration over $S^1$ in the $Q$-flux model, the coordinates become non-commutative and the appropriate
mathematical structure is that of a non-commutative 2-torus fibred over $S^1$. We may go a step further
and perform yet another T-duality by thinking of the original flux background as $T^3$ fibration
over a point (it would be $M$ for more general base manifolds). The resulting closed string
background (hereafter called $R$-flux model) not only fails to be globally geometric, but also locally.

One may object the validity of duality rules in the $R$-flux model, which are based on the transformation
of the 2-form local field $B$, because $B$ cannot be coordinate independent on
$T^3$ that supports the 3-form $H = dB$. Note, however, that this issue does not invalidate the analysis, as can be seen
in the context of conformal field theory, but it is intimately related to the absence of
geometric structures even locally. Then, because of all these, the coordinates of the $R$-flux model
become not only non-commutative but also non-associative and one talks of a non-associative 3-torus
fibred over a point. Clearly, the picture persists for higher dimensional toroidal fibrations with fluxes.
One obtains similar qualitative features for fibrations over more general base manifolds $M$, though the details may
slightly differ from case to case.

One can take the attitude to ignore all these new mathematical structures, as exotica, and stick
to the original formulation of flux backgrounds in terms of ordinary geometry. However, as long as
T-duality is used as building principle in string theory, geometric as well as non-geometric backgrounds
\cite{Kawai:1986va,Lerche:1986cx,Antoniadis:1986rn,Narain:1986qm,Dabholkar:2002sy,Hellerman:2002ax,Flournoy:2004vn,Hull:2004in,
Shelton:2005cf,Dabholkar:2005ve,sethi}
should be accounted equally and, for that matter, non-commutative and even non-associative
spaces should be on par with differentiable manifolds.
The ultimate formulation of the theory simply cannot do without them. Moreover, there are generic non-geometric
string vacua that cannot be T-dualized to any geometric space.  A closely related question
is the possibility to obtain an alternative understanding of (some of) the coordinate dependent
compactifications, also called Scherk-Schwarz reductions  \cite{Scherk:1978ta}, that lead to gauged supergravities and
generate mass terms in lower dimensions. They might also admit an equivalent description as reductions
on non-commutative and/or non-associative spaces when formulated in terms of some suitably chosen T-dual
non-geometric backgrounds that are insensitive to geometric concepts, such as local coordinate dependence,
exactly as for the 2-form B-fields on 3-tori that were discussed above. Conversely, one may ask whether
non-commutative and/or non-associative spaces with non-geometric fluxes can be used to uplift lower dimensional gauged supergravity
models in string and M-theory, thus filling a gap in the literature. This was recently demonstrated
for non-geometric $Q$- and $R$-flux backgrounds, which allow a description in terms of freely acting asymmetric orbifold spaces \cite{Condeescu:2013yma}.
Other problems of modern day
string theory may also find an intrinsic solution through these new mathematical structures.

At the core of the problems lies the absence of
a systematic formulation of (globally or even locally) non-geometric backgrounds
of string theory, using a duality invariant framework. While many more things remain to be understood
and be done in this direction, hopefully through the development and proper use of T-folds \cite{Hull:2004in} as substitute to
ordinary manifolds, we build on existing techniques and exploit them to gain better understanding of
non-geometry for the simplest case of $T^3$ torus fibrations with fluxes over a point. A prime
directive of the programme is to provide an intuitive and user friendly guide on how to deal with non-commutative and
non-associative generalizations of geometry. We report partial progress on this problem by considering
the duality invariant framework of double phase space, in which all coordinates, momenta and their dual counterparts
appear on equal footing, and construct a star-product on the space of its functions that is applicable to
the T-dual faces of the toroidal flux model. Our results are briefly compared to other that appeared in the
literature before, most notably in \cite{Mylonas:2012pg}, using a somewhat different stand point.

A particularly useful framework to explore non-commutativity/non-associativity is provided by the
magnetic field analogue of non-geometric flux models, extending some well understood structures in physics
in a natural way. It has been known for a long time that an electrically charged point-particle moving in
the background of magnetic field exhibits non-commutativity among its momenta. There is no violation of
Jacobi identity as long as the magnetic field satisfies Maxwell's equations. Quantum mechanics is well defined,
using operators acting on Hilbert space, and proper treatment of the problem gives rise to the well known
theory of Landau levels. Generalizing this picture to Dirac's variant of Maxwell theory by adding sources of
magnetic charge, such as monopoles or other distributions of magnetic charge in space, is not that simple, in general.
The classical description of the point-particle dynamics becomes non-associative and for that reason the standard rules of
canonical quantization cannot be used anymore. The investigation of this problem turns out to be very illuminating, indeed,
and the results can be easily taken to the phase space of non-geometric flux models by exchanging the role of coordinates
and momenta (modulo the periodicity in the toroidal directions).

In section 2, we present a brief outline of the so called parabolic flux models that arise by successive application of
T-duality on toroidal flux backgrounds. The non-geometric faces of these models provide the framework to explore
non-commutativity and non-associativity in closed string theory in simplest possible terms.  In section 3, we use
the cohomology theory of Lie algebras and Lie groups to characterize the violation of Jaboci identity in non-associative
geometry by 3-cocycles. Different cohomology theories are used to provide complementary descriptions of the obstruction.
In section 4, we introduce a non-associative star-product on functions in the phase space of non-geometric vacua and extend
it to double phase space in a duality invariant way. The resulting algebraic structure defines
non-commutative and non-associative tori in closed string theory and substitutes for canonical quantization.
In section 5, we discuss analogies with the quantization of charged particles in the background of a
Dirac monopole, and generalizations thereof, where 3-cocycles also arise as obstructions to the Jacobi identity.
Several physical questions are also addressed in this context, trying to understand the implications of
non-associativity in more elementary terms. Finally, in section 6, we present the conclusions and discuss some open problems.
There is also an appendix summarizing the cohomology theory of Lie algebras and Lie groups.

\section{T-dual faces of toroidal flux backgrounds}
\setcounter{equation}{0}

The constant flux backgrounds in three dimensions are the simplest examples of geometric and non-geometric spaces
that coexist in closed string theory.
They come in four different versions, which are related to each other by successive T-duality transformations. Here, we
give a brief description of their occurrence and their mathematical properties, following earlier work on the subject
\cite{Lust:2010iy, Condeescu:2012sp, Andriot:2012vb} and \cite{Blumenhagen:2010hj, Blumenhagen:2011ph} to which we also
refer the reader for more details. This provides the main framework for the present work.

More concretely, we consider the compactification of string theory on a three-dimensional
torus $T^3$, which can be viewed as ${\cal F}=T^{n}$ fibration  over a $(3-n)$-dimensional base
${\cal B}=T^{3-n}$, and examine four different cases, letting $n=0,1,2,3$.
The starting point, corresponding to $n=0$, is a flat three-torus $T^3$ with coordinates $x^i$, $i=1,2,3$, and periodic
identifications along the cycles $x^i \sim x^i + 2 \pi r^i$, where $r^i$ denote the three radii of $T^3$. There is also
an $H$-flux 3-form in the model,
\be
H_3=H ~ d X^1 \wedge d X^2 \wedge d X^3 ~,
\ee
where $H$ is constant. It has to obey a quantization condition, which is of topological origin and it is given by the
integral formula
\begin{equation}
{1\over 4\pi^2}\int H=k\, ,\quad k\in{\mathbb Z}\, .
\end{equation}
In practice, it suffices to concentrate on the simplest case with topological charge $k=1$.

The components of the anti-symmetric tensor field $B_{ij}$ are fixed to a particular gauge, so that $H_3 = dB$.
Different choices are related to each other by $B^{\prime} = B + d \Lambda$, where $\Lambda$ is an arbitrary
$1$-form. Let us first make the choice
\be
B_{12} = H x^3 ~, ~~~~~~ B_{23} = 0 =  B_{31} ~,
\label{lesssymme}
\ee
which is the easiest to work, as it simplifies the presentation. Soon afterwards, we will make another choice
for $B$ that is more symmetric and stick with it in the remainder.

T-duality transformations act on both metric and anti-symmetric tensor fields and they are typically described by
Buscher rules \cite{Buscher:1987sk} in the presence of isometries. There is also a canonical formulation of T-duality
based on phase space techniques that provides a systematic definition of dual coordinates and dual conjugate momenta
\cite{Alvarez:1994wj}.
In all cases, we have the following commutation relations among the coordinates $x^i$ of the original torus and their
conjugate momenta $p^i$,
\be
[x^i , ~ p^j] = i \delta^{ij} ~, ~~~~~~
[p^i , ~ p^j] = 0 ~.
\ee
Likewise, the commutation relations among the would be dual coordinates $\tilde{x}^i$ of the model and their conjugate
momenta $\tilde{p}^i$ are
\be
[\tilde{x}^i , ~ \tilde{p}^j] = i \delta^{ij} ~, ~~~~~~
[\tilde{p}^i , ~ \tilde{p}^j] = 0 ~.
\ee
What differentiates the four different T-dual faces of the toroidal flux model are the commutation relations among the
coordinates $x^i$ and their dual counterparts $\tilde{x}^i$, which are non-trivial. This is precisely what makes the theory
non-commutative and non-associative in the presence of fluxes.

The situation is more intricate than the old problem regarding the fate of isometry groups, and in some cases of supersymmetry,
after duality. It has been pointed out that isometries with fixed points, e.g., rotational 
isometries, give rise to coordinate 
dependent compactifications of the fermionic sector in that the Killing coordinates enter into the
definition of supersymmetry transformations and Killing spinor equations. After T-duality, space-time supersymmetry 
appears to be lost in the effective theory, whereas the world-sheet supersymmetry is realized non-locally through the 
dual Killing coordinate \cite{Bakas:1995hc}. The present work can be regarded as continuation of
those old observations inspired by the new developments in the subject.

Let us now briefly explain how exactly non-commutativity/non-associativity come into play, summarizing the non-trivial
relations among the coordinates and their dual that arise in the frame \eqn{lesssymme}, where the calculations become simpler.
Here, we adopt the dilute flux approximation, i.e., $H/{\rm Vol}(T^3) << 1$, so that the flux background is 
an approximate fixed point of the renormalization group equations and the computations are performed to linear order in $H$ 
in the framework of the conformal field theory $CFT_H$, as explained in the original work \cite{Andriot:2012vb, Blumenhagen:2011ph} 
to which we refer for all technical details. Then, in this context, T-duality has a well defined world-sheet description in terms 
of chiral currents, all the way, and it will be used without explanation to state the results. An alternative understanding 
of the same problem based on canonical methods will appear elsewhere \cite{baklus}. 

We have, in particular, the following four cases:

{\bf H-flux model}: This is the original geometric model of a 3-torus with constant $H$-flux that corresponds to $n=0$.
Obviously, the coordinates $x^i$ commute among themselves and they also commute with the dual coordinates $\tilde{x}^i$
that are in our disposal, but do not participate in the field theory description of the original geometric face of the model.
Nevertheless, computing the commutation relations among the would be dual coordinates of the model , one finds that
$[\tilde{x}^1, ~ \tilde{x}^2] \sim H \tilde{p}_3$, whereas all other commutators vanish in the frame \eqn{lesssymme}.

{\bf f-flux model}: Performing a T-duality transformation along the one-dimensional circle fibre ${\cal F}=T^{1}_{x_1}$ in the
$x_1$-direction, one obtains the Heisenberg nilmanifold, which is a twisted torus without $B$-field. It corresponds to the
case $n=1$ and it is topologically distinct from $T^3$. In geometrical terms, the original 3-torus is Bianchi-I, whereas
the twisted torus is Bianchi-II. Then, one finds that the coordinates $x^i$ commute among themselves and the same holds
for the dual coordinates $\tilde{x}^i$. There are, however, non-trivial commutation relations among $x^i$ and $\tilde{x}^i$
which take the form $[x^1, ~ \tilde{x}^2] \sim f \tilde{p}^3$, whereas all other commutators vanish in the frame
\eqn{lesssymme}. The constant $f$ is the same as $H$, but here we are using a different symbol to indicate explicitly
that we are referring to this particular T-dual face of the original flux model.

{\bf Q-flux model}: Performing a T-duality transformation in the $x_1$ and $x_2$ directions on the two-dimensional torus fibre
${\cal F}=T^{2}_{x_1,x_2}$, we obtain the next model in the series, corresponding to $n=2$.
This new background is again a $T^2$-fibration, but the corresponding metric and $B$-field are defined only locally and
not globally. The reason for failing to be a Riemannian manifold is provided by the fact
that the fibre ${\cal F}$ has to be glued together by a T-duality transformation when transporting it once around the base
${\cal B}$ and not by a standard diffeomorphism. Here, one finds the non-trivial commutation relations
$[x^1, ~ x^2] \sim Q \tilde{p}^3$, whereas all other commutators of the coordinates are zero in the frame \eqn{lesssymme}.
All dual coordinates commute among themselves and they also commute with $x^i$. As before, we are using a different
name for the constant flux, this time calling it $Q$ to distinguish this particular face of the model from the other.

{\bf R-flux model}: Finally one can consider a T-duality transformation along the entire three-dimensional torus, which is
seen as a fibration over a point with fibre ${\cal F}=T^{3}_{x_1,x_2,x_3}$. It corresponds to the last case, $n=3$.
Note, however, that this T-duality looks problematic, since the $x_3$ direction is not any longer a Killing isometry of
the space as $B_{ij}$ depends explicitly on it. Hence, the standard Buscher rules cannot be applied as they stand.
Nevertheless, in the context of conformal field theory, the prescription how to proceed is well defined. One performs the
final T-duality by flipping the sign of the corresponding coordinate and ends up with a "space" that is left-right
asymmetric in all its three directions. Then, there are non-trivial commutation relation among the coordinates,
which take the form $[x^1, ~ x^2] \sim R p^3$. The other commutators vanish in the frame \eqn{lesssymme}. As for the
dual coordinates $\tilde{x}^i$, it turns out that they commute among themselves as well as with the original coordinates
$x^i$. Again, to distinguish this face of the model from the other, we denote the constant flux by $R$. The $R$-flux
model is non-geometric locally as well globally. It is the exact opposite face of the $H$-flux model in that
it can be solely described in terms of the dual coordinates, whereas the original toroidal coordinates are spectators.

Repeated use of T-dualities brings to light the algebraic structure of the dual coordinates, making them part of
"space" for closed string theory.
The monodromies of the toroidal fibrations play important role in each step of the way that led to the novel
commutation relations among the toroidal coordinates and/or their dual. The monodromies
specify the gluing conditions of the fibre when going around the base, telling how the complex structure and the
complexified K\"ahler class are transforming.
As explained in \cite{Andriot:2012vb}, these flux models are called parabolic because the monodromies that define the
backgrounds in each T-dual face are of infinite order, i.e., parabolic.

There is another choice of $B$-field that is linear and more symmetric in the coordinates, compared to the less
symmetric choice \eqn{lesssymme}, namely
\be
B_{12} = {H \over 3} x^3 ~, ~~~~~~ B_{23} = {H \over 3} x^1 ~, ~~~~~~ B_{31} = {H \over 3} x^2 ~,
\label{newframei}
\ee
which is obtained by a gauge transformation of the 2-form $B$-field, $ B \rightarrow B + d \Lambda$, letting  
$\Lambda = (x_2 x_3 dx_1 + 3 x_1 x_3 dx_2 + 2 x_1 x_2 dx_3)/2$.
In this case, all components of the $B$-field depend on the toroidal coordinates and one has to resort to
conformal field theory techniques, and not just Buscher rules, to implement the dualities at each step.
In this symmetric frame, non-geometry makes its appearance from the very beginning, but we are going to keep
the same name for the different T-dual faces of the toroidal flux model. Then, in this new frame, the
commutation relations among the coordinates and their dual take the following form in each T-dual face:
\ba
{\bf H}: & & ~~ [x^i , ~ x^j] = 0 ~, ~~~~~ [x^i , ~ \tilde{x}^j] = 0 ~, ~~~~~
[\tilde{x}^i , ~ \tilde{x}^j] = i H \epsilon^{ijk} \tilde{p}^k ~, \\
{\bf f}: & & ~~ [x^i , ~ x^j] = 0 ~, ~~~~~ [\tilde{x}^i , ~ \tilde{x}^j] = 0 ~, ~~~~~
[x^i , ~ \tilde{x}^j] = i f \epsilon^{ijk} \tilde{p}^k ~, \\
{\bf Q}: & & ~~ [\tilde{x}^i , ~ \tilde{x}^j] = 0 ~, ~~~~~ [x^i , ~ \tilde{x}^j] = 0 ~, ~~~~~
[x^i , ~ x^j] = i Q \epsilon^{ijk} \tilde{p}^k ~, \\
{\bf R}: & & ~~ [\tilde{x}^i , ~ \tilde{x}^j] = 0 ~, ~~~~~ [x^i , ~ \tilde{x}^j] = 0 ~, ~~~~~
[x^i , ~ x^j] = i R \epsilon^{ijk} p^k
\ea
In writing these results we have absorbed all factors arising in the calculations into the constant flux coefficient
$F = H, ~ f, ~ Q$ or $R$ (which one depends on the T-dual face)
and kept the same name for notational convenience. After all, such factors are not important for the purposes of the
present work.

From now on, we stick with the symmetric choice \eqn{newframei} for the $B$-field and use it in the
following\footnote{We note at this end that there is an integral formula for writing the commutator among the
coordinates (and their dual counterparts) in a frame invariant way. It involves a contour integral of the
3-form flux along certain cycles of the model and makes sense in all $B$-frames \cite{Andriot:2012vb}.}.
In Table 1, we summarize the results for the four T-dual parabolic flux backgrounds, for later use, listing
only the non-trivial commutation relations between the coordinates $x^i$ and their dual counterparts  $\tilde{x}^i$.
The fluxes introduce non-commutativity as well as non-associativity in string theory, as can be seen by
computing the associator of the coordinates.


\begin{table}[h]
\centering
\renewcommand{\arraystretch}{1.3}
\tabcolsep10pt
\begin{tabular}{|c||c|c|}
\hline
T-dual frames  & Commutators & Three-brackets \\  \hline\hline
$H$-flux & $[\tilde{x}^i, ~ \tilde{x}^j] = i H \epsilon^{ijk} \tilde{p}^k$ & $[\tilde{x}^1, ~ \tilde{x}^2, ~ \tilde{x}^3] \sim H$\\
$f$-flux & $[x^i, ~ \tilde{x}^j] = i f \epsilon^{ijk} \tilde{p}^k $ & $[ x^1, ~ \tilde{x}^2, ~ \tilde{x}^3] \sim f$\\
$Q$-flux & $[ x^i, ~ x^j] = i Q \epsilon^{ijk} \tilde{p}^k$ & $[x^1, ~ x^2, ~ \tilde{x}^3] \sim Q$ \\
$R$-flux & $[ x^i, ~ x^j] = i R \epsilon^{ijk} p^k$ & $[ x^1, ~ x^2, ~ x^3] \sim R$ \\ \hline
\end{tabular}
\caption{\small  Non-vanishing commutators and three-brackets in the parabolic flux backgrounds.
\label{tablemomwind2} }
\end{table}

One may use the commutation relations of these models to define what is often called a twisted Poisson
structure\footnote{Early work on the subject includes \cite{sereva}, whereas recently this algebra was described by
quantizing 2-plectic manifolds in loop space using
groupoids \cite{Saemann:2012ex}. We are not making use of this terminology here.}, which is a rather new object in physics.
The emergence of this mathematical structure is a stringy feature, related
to the fact that the closed string can move in non-geometric spaces. Often, but not always, T-duality
relates left-right symmetric closed string backgrounds to left-right asymmetric ones and we note here that
the non-commutative and, in particular, the non-associative geometries refer to the coordinates of the left-right asymmetric spaces.
The $R$-flux background appearing in the list of our models is left-right asymmetric in all three string coordinates.

There is an alternative systematic derivation of all these commutation relations, using the canonical approach to
T-duality transformations. The method applies to the parabolic model, but it also generalizes nicely to other non-geometric
backgrounds. They include the so called elliptic and double elliptic flux models \cite{Lust:2010iy, Condeescu:2012sp},
whose canonical formulation requires making use of the full double field theory phase space of coordinates, momenta, and their
dual and they exhibit some more intricate mathematical structures. It is beyond the scope of the present work to delve into
those generalizations. Further details will appear elsewhere \cite{baklus}.

\section{Cohomology of the parabolic flux model}
\setcounter{equation}{0}

The non-geometric spaces arising in closed string theory exhibit some novel features that we will try to understand
better using phase space techniques and star-products. In view of this formulation, and in order to obtain a
precise mathematical characterization of the violation of Jacobi identity, we are going to employ
the cohomology theory of Lie algebras and Lie groups.
The formalism will be developed to the level of double field theory phase space, which is most appropriate for
stating our results in a manifestly T-duality invariant way.

\subsection{Deformations of Lie algebras}

The commutation relations among the coordinates and momenta of the parabolic flux models give rise to some
new mathematical structures that are non-trivial deformations of the standard relations one usual gets
in classical as well as quantum mechanics. The deformation theory of Lie algebras is a well developed subject
and its proper mathematical formulation relies on cohomology.

To motivate the presentation, we introduce by simple dimensional analysis
the two physical constants that were silently normalized to $1$ in the previous discussion, namely Planck's constant
$\hbar$ and the string tension $T$. We have $T= (2 \pi \alpha^{\prime} \hbar)^{-1}$, 
whereas the string length is given by $l_s = \sqrt{\hbar / T}$ in units where the speed of light is $c=1$. 
The flux has units of inverse length so that the $B$-field is dimensionless like the metric. 
Then, for the parabolic models we have schematically the non-trivial relations
\begin{equation}
[ x^i, ~ x^j ] \sim  {\hbar \over T^2} F \epsilon^{ijk} p_k ~,
\label{commclosed}
\end{equation}
whereas the remaining commutation relations assume their standard form,
\be
[x^i, ~ p^j]=i \hbar \delta^{ij}~, ~~~~~ [p^i, ~ p^j]=0 ~.
\label{commclo23}
\ee
Depending on the given duality frame, the flux $F$ appearing in equation \eqn{commclosed} corresponds to either
$F=H, ~ f, ~ Q$ or $R$-flux, respectively.
Also, the variables $x^i$ and $x^j$ denote collectively the coordinates or dual coordinates of the 3-torus,
depending on the chosen duality frame, and, likewise, $p^k$ denote the momentum or the dual momentum along the
toroidal directions. To avoid confusion, one may stick with the $R$-flux model, thinking of $x^i$ and $p^j$
as its coordinates and momenta, and also impose the usual commutation relations among the dual variables,
\be
[\tilde{x}^i, ~ \tilde{x}^j]=0 ~, ~~~~~ [\tilde{x}^i, ~ \tilde{p}^j]=i \hbar \delta^{ij}~, ~~~~~ [\tilde{p}^i, ~ \tilde{p}^j]=0 ~,
\ee
assuming that all other commutators among tilded and/or untilded variables vanish. The other T-dual faces of
the toroidal flux model can be described analogously, using Table 1 as a guide for the necessary relabeling of tilded and
untilded variables.

The most important algebraic result is that the triple bracket among the closed string coordinates or their dual counterparts
(depending on the chosen duality frame) turns out to be non-vanishing. We have schematically
\begin{equation}
[ x^1, ~ x^2, ~ x^3]:=[[ x^1, ~ x^2], ~ x^3]+{\rm cycl.~perm.} \sim  {\hbar^2 \over T^2}  F\, ,
\label{asso}
\end{equation}
which demonstrates vividly not only the non-commutative, but the non-associative aspects of all $H, ~ f, ~ Q, ~ R$-deformed
closed string models as well. Actually, even if we were considering the $H$-flux model, which is purely geometric in space,
the non-associative structure would reside in the dual space and come to light only after successive application of
T-duality transformations until one reaches the purely non-geometric $R$-flux model. Proper treatment of the problem
requires the use of double phase space and, thus, non-associativity becomes inevitable.
We call the triple bracket $[ x^1, ~ x^2, ~ x^3]$ the {\em associator} of the three coordinates, but equally
well it can be called Jacobiator, as it describes the deviation from having a Lie algebra structure.

The loss of associativity is a combined effect that involves Planck's constant as well as the flux,  
as can be seen by inspecting the right-hand side of equation \eqn{asso} that is proportional to $\hbar^2 F / T^2$.
When $\hbar = 0$, all generators commute among themselves and the commutation relations \eqn{commclosed} and \eqn{commclo23} 
can be regarded mathematically as deformation of
the Abelian algebra of translations ${\bf t}_6$ acting on six-dimensional phase space $(x^i, p^i)$. 
On the other hand, by turning on $\hbar$ while keeping $F$ zero we obtain the Heisenberg algebra ${\bf g}$ for the three
toroidal coordinates and their conjugate momenta, which is a central extension of the Abelian group of translations in the
corresponding six-dimensional phase space.  Thus, the complete set of
commutation relations \eqn{commclosed} and \eqn{commclo23}, when both parameters are turned on, correspond to a 
deformation of any one of the two intermediate Lie algebras, ${\bf t}_6$ or ${\bf g}$.

In Table 2, we summarize the different limiting cases of the commutation relations \eqn{commclosed} and \eqn{commclo23}, 
setting the physical deformation parameters $\hbar$ or $F$ equal to zero, and identify
the resulting Lie algebras that will be used in the following.

\begin{table}[h]
\centering
\renewcommand{\arraystretch}{1.3}
\tabcolsep10pt
\begin{tabular}{|c||c|c|}
\hline
Contraction & Structure of commutators & Name of Lie algebra \\  \hline\hline
$\hbar = 0$, ~ $F$ \, {\rm any} & $[x^i, x^j] = 0$, ~ $[x^i, p^j] = 0$ & Algebra of translations ${\bf t}_6$ \\
$\hbar \neq 0$, ~ $F=0$ & $[x^i, x^j] = 0$, ~ $[x^i, p^j] = i \hbar \delta^{ij}$ & Heisenberg algebra {\bf g} \\ \hline
\end{tabular}
\caption{\small The different contractions of the parabolic flux model; in all case, $[p^i, p^j] = 0$.
\label{tablemomwind2} }
\end{table}

Next, we adopt the deformation approach to the problem and make it quantitative using the cohomology theory of the Abelian algebra
of translations (with $\hbar = 0$) and that of the Heisenberg algebra (with $F = 0$). The two algebraic frameworks are
complementary to each other, although the characterization of the associator differs in Lie algebra cohomology. 
With these explanations in mind, we set from now on the two physical constants $\hbar$ and $T$ back to their normalized values
$1$ and will reinstate them in the text only when it is necessary.

\subsection{3-cocycles in Lie algebra cohomology}

We outline the two alternative cohomological interpretations of the commutation
relations \eqn{commclosed}, \eqn{commclo23} and the associator \eqn{asso} assigned to the non-geometry of the
parabolic flux models. The results will also be used later to guide the construction of star-products of functions
in phase space. Here, we focus on the
cohomology of the $R$-flux model, in the notation of Table 1, but the same treatment applies to the algebra of
the dual coordinates and  momenta of the $H$-flux model, trading all untilded variables with the tilded ones.
At the end of this subsection we will extend the cohomological description to the double phase space of the parabolic
flux models, encompassing all coordinates, momenta and their dual counterparts on equal footing, and provide a
duality covariant formulation of Lie algebra cohomology that also accounts for the structure of the $f$- and $Q$-flux models.
We refer the reader to the appendix for the appropriate definitions and terminology.

The first interpretation relies on the Abelian algebra of translations in phase space with generators $T_I = x^i, ~ p^i$
arising in the contraction limit $\hbar = 0$ irrespective of flux. The cohomology groups of the algebra of translations are non-trivial
when the cochains have real-valued coefficients. Choosing, in particular, a 3-cochain with $c_3(x^1, x^2, x^3) = 1$, up to
normalization, and $c_3(T_I, T_J, T_K) = 0$ for all other choices of generators (when at least one of the $T$'s is a momentum
generator), we see that $c_3$ satisfies the 3-cocycle condition $dc_3 (T_I, T_J, T_K, T_L) = 0$, namely
\ba
& & c_3([T_I , T_J], ~ T_K, ~ T_L) - c_3([T_I , T_K], ~ T_J, ~ T_L) +
c_3([T_I , T_L], ~ T_J, ~ T_K)  + \nonumber\\
& & c_3([T_J , T_K], ~ T_I, ~ T_L) - c_3([T_J , T_L], ~ T_I, ~ T_K) +
c_3([T_K , T_L], ~ T_I, ~ T_J) = 0 ~.
\ea
The result follows from the simple fact that $[T_I , ~ T_J] = 0$ for all generators of the algebra of translations in phase space.
Clearly, $c_3$ is a genuine 3-cocycle and not a coboundary. Otherwise, it would take the form  $c_3(T_I , T_J,  T_K) =
\theta ([T_I , T_J], ~ T_K) + \theta ([T_J , T_K], ~ T_I) + \theta ([T_K , T_I], ~ T_J)$ for an appropriately chosen
real-valued 2-cochain $\theta (T_I, T_J)$. But this is impossible by the Abelian nature of the algebra, unless, of course, $c_3$
vanishes identically. Thus, in this description,
the associator has the interpretation of a non-trivial 3-cocycle of the Abelian algebra of translations in phase space with
real-valued coefficients in cohomology. We write
\be
[x^1, ~ x^2 , ~ x^3] \sim c_3 (x^1, x^2, x^3) ~.
\label{tsiaoua3}
\ee
The 3-cocycle of the parabolic flux model has support only in the toroidal directions $x^i$ and it is constant
everywhere\footnote{It should be compared to the 3-cocycle of the algebra of translations that arises in a Dirac
monopole field and obstructs the Jacobi identity, but it has support only at a point - the location of the pole.
Detailed comparison of the two models will be made later in section 5. We only note here that the analogue of 
flux in closed string models is provided by the magnetic charge in monopole backgrounds.}.

The second cohomological description of the associator \eqn{asso} views the commutation relations \eqn{commclosed}
and \eqn{commclo23} as a particular deformation of the Heisenberg algebra generated by $x^i$, $p^i$ and $\mathbb{1}$
by a 2-cochain $c_2$ taking values in the Heisenberg algebra itself. The cochain is chosen so that
$c_2(x^i , x^j) = \epsilon^{ijk} p^k$, up to a multiplicative constant, and it vanishes for any other pair of generators
apart from the $x^i$'s.
Any 2-cochain is by definition an anti-symmetric bilinear map and, therefore, $c_2$ introduces the necessary deformation term
$[x^i ~, x^j] = i c_2(x^i, x^j) \sim i\epsilon^{ijk} p^k$ when added to the commutation relations
of the Heisenberg algebra. Then, the associator \eqn{asso} can be described in terms of Lie algebra cohomology for the Heisenberg
algebra ${\bf g}$ with coefficients in ${\bf g}$. The action of the
coboundary operator $d$ on a 2-cochain $c_2 \in C^2 ({\bf g} , {\bf g})$ takes the following form in Chevalley-Eilenberg cohomology,
\ba
dc_2 (x^1 , x^2, x^3) & = & -c_2([x^1, ~ x^2] , ~ x^3) + c_2([x^1 , ~ x^3] , ~ x^2) - c_2([x^2 , ~ x^3] , ~ x^1) \nonumber\\
& & + \pi(x^1) c_2(x^2 , x^3) - \pi(x^2) c_2(x^1 , x^3) + \pi(x^3) c_2(x^1 , x^2)
\ea
using the adjoint representation of the Heisenberg algebra ${\bf g}$ on ${\bf g}$, i.e., $\pi({\bf g}) =
{\rm Ad}_{\bf g} = [{\bf g} , ~ \cdot ~]$, which accounts for the terms appearing in the second line.
All terms in the first line vanish identically, since the $x^i$'s commute among themselves in ${\bf g}$. Thus, in this case,
only the terms in the second line contribute to the answer,
\be
dc_2 (x^1 , x^2, x^3) = [x^1 , ~ c_2(x^2 , x^3)] - [x^2 , ~ c_2(x^1 , x^3)] + [x^3 , ~ c_2(x^1 , x^2)] ~.
\ee
The right-hand side is nothing else but the associator $[x^1, ~ x^2 , ~ x^3]$, up to a sign, for the particular choice of
the 2-cochain $c_2$ made above. We also note that $dc_2$ vanishes when computed for any other triplet of elements in ${\bf g}$.
Thus, we arrive at the final result
\be
[x^1, ~ x^2 , ~ x^3] \sim dc_2 (x^1, x^2, x^3) ~,
\label{motre1}
\ee
showing that the associator differs from zero by an exact 3-cocycle term in the appropriate Chevalley-Eilenberg cohomology
for the Heisenberg algebra ${\bf g}$. Note, in this respect, that $dc_2 (x^1, x^2, x^3)$ is real-valued and, hence, it is
proportional to the generator $\mathbb{1}$ that belongs in ${\bf g}$ like the other generators $x^i$ and $p^i$. If $c_2$
were a 2-cocycle, satisfying the special $dc_2 = 0$, associativity would be fully restored, but, of course, this is note the case
here.

The difference between the two interpretations of \eqn{asso} arises because a trivial cocycle in one cohomology might not
be trivial in another. The interpretation of the associator as an exact 3-cochain in the cohomology theory of the Heisenberg algebra
${\bf g}$ with coefficients in ${\bf g}$ does not imply that the obstruction can be removed. It rather means that there is no
obstruction to integrating such infinitesimal deformations of the algebra (recall that the third cohomology group
$H^3 ({\bf g} , {\bf g})$ describes the obstructions to integrating infinitesimal deformations of the algebra and we are
lucky that $dc_2$ is a trivial element of it). The second interpretation of associator as \eqn{motre1} may also help to uncover
relations with homotopy algebras. Further exploration of this idea is left open to future work.
In any case, the violation of Jacobi identity is an obstruction to assigning operators to the generators, which could then
act in a common domain of a Hilbert space, as in the conventional formulation of quantum mechanics.

Actually, what we have described so far is only half of the story, since the parabolic flux model has additional phase space
variables $(\tilde{x}^i, \tilde{p}^i)$ satisfying their own commutation relations. In the $R$-flux face, the tilded generators
form a second Heisenberg algebra, without any deformation terms, and they commute with the untilded generators. Thus, the
complete cohomological interpretation of the $R$-flux model can be given either in terms of the Abelian algebra of
translations in double phase space or in terms of the algebra ${\bf g} \oplus {\bf \tilde{g}}$ formed by the direct sum
of two copies of the Heisenberg algebra (one for the untilded and one for the tilded phase space variables).

The first description is based on real-valued cohomology of the Lie algebra of translations in
double phase space. The 3-cocycle is taken to be $c_3 (x^1 , x^2 , x^3) = 1$, up to
normalization, and it is zero for any other choice of the three generators (tilded or untilded).
Then, the obstruction to Jacobi identity is given by $c_3$, as in equation \eqn{tsiaoua3}. In the second description,
we consider the cohomology of the Lie algebra ${\bf g} \oplus {\bf \tilde{g}}$ with coefficients in
${\bf g} \oplus {\bf \tilde{g}}$. We introduce a 2-cochain $c_2$ that vanishes for all other choices of generators
(tilded or untilded) apart from $c_2 (x^i , x^j) = \epsilon^{ijk} p^k$, up to a multiplicative constant, and note that
the obstruction to Jacobi identity is given by $dc_2$, as in equation \eqn{motre1}. Either way, the extension of
Lie algebra cohomology to the double phase space of the $R$-flux model appears to
be cosmetic, since the tilded coordinates act as spectators. The very same statements can be made for the $H$-flux face
of the parabolic models, provided that the tilded and untilded variables are interchanged everywhere.

The double phase space is advantageous for the unified description of all T-dual faces of the toroidal flux models.
The algebraic deformations of the $f$- and $Q$-flux models are also described in terms of the cohomology
of the Lie algebra of translations in double phase space with real valued coefficients or equivalently in terms
of the cohomology of the Lie algebra ${\bf g} \oplus {\bf \tilde{g}}$ with coefficients in ${\bf g} \oplus {\bf \tilde{g}}$.
The only difference lies in the support of the corresponding cochains, which require the use of both tilded and untilded
generators. For the $f$-flux model, the first cohomological description is based on the choice of a 3-cocycle that
vanishes for all other choices of generators apart from $c_3(x^i , \tilde{x}^j , \tilde{x}^k) \sim \epsilon^{ijk}$.
The second cohomological description of the $f$-flux model is based on the choice of a 2-cochain $c_2$ that vanishes for
all other choices of generators apart from $c_2 (x^i , \tilde{x}^j) \sim \epsilon^{ijk} \tilde{p}^k$. Either way, the
cohomological interpretation of the associators parallels that of the $R$- or $H$-flux model by introducing tilded and
untilded variables in the appropriate places.
Likewise, there are two cohomological descriptions of the $Q$-flux model based on the choices
$c_3(x^i , x^j , \tilde{x}^k) \sim \epsilon^{ijk}$ and $c_2 (x^i , x^j) \sim \epsilon^{ijk} \tilde{p}^k$,
respectively. Table 1 is a good guide for the choices one has to make in each case separately.

Thus, in effect, we have obtained a unified cohomological description of the algebraic deformations introduced by dualities
on the double phase space of the parabolic flux models.

\subsection{3-cocycles in Lie group cohomology}

Next, we reformulate the deformations at the group level, using the associated theory of Lie
group cohomology. The result will be used in the next section to justify the construction of the star-product
in phase space. For definiteness, we choose to work with the $R$-flux model and discuss only at the very end the
extension of the formalism to the other dual faces of the parabolic flux model.

Let us exponentiate the action of the position and momentum generators and consider the group elements (formal loops)
\be
U(\vec{a}, ~ \vec{b}) = e^{i(\vec{a} \cdot \vec{x} + \vec{b} \cdot \vec{p})} ~.
\label{basiako}
\ee
They satisfy the following product relation, based on the Baker-Campbell-Hausdorff formula,
\be
U(\vec{a}_1, ~ \vec{b}_1) U(\vec{a}_2, ~ \vec{b}_2) =
e^{-{i \over 2} (\vec{a}_1 \cdot \vec{b}_2 - \vec{a}_2 \cdot \vec{b}_1)}
U \left(\vec{a}_1 + \vec{a}_2, ~~ \vec{b}_1 + \vec{b}_2 - {R \over 2} (\vec{a}_1 \times \vec{a}_2) \right) .
\label{groupcompos}
\ee
Note in passing that the validity of
Baker-Campbell-Hausdorff formula is questionable when the Jacobi identity is violated and this is closely related to the
ambiguities in defining the exponential of a non-associative algebra by power series. An analogous formula for non-associative
algebras has appeared in the literature based on a particular definition of the exponential function so that
$e^A e^A = e^{2A}$ \cite{lukas1} (but see also \cite{lukas2} for an update of the recent developments in the subject). We
are not going to worry about it here\footnote{In other non-geometric string backgrounds the algebraic deformations can be more
complex and all issues raised above need to be addressed properly.}, because any deviations from
$e^A e^B = {\rm exp} (A + B + [A, ~ B]/2 + \cdots)$ 
appear at the level of triple commutators or higher, which are not contributing to \eqn{groupcompos}. Besides, we already know
from the discussion of section 3.2 that there is no obstruction to integrating the Lie algebra deformations arising in the
parabolic model.

If $R$ were zero, equation \eqn{groupcompos} would simply be the defining relation of a projective representation of the
Abelian group of translations in phase space driven by the real-valued 2-cocycle
\be
\varphi_2 (\vec{a}_1 , \vec{b}_1 ; \vec{a}_2 , \vec{b}_2) = \vec{a}_1 \cdot \vec{b}_2 - \vec{a}_2 \cdot \vec{b}_1 ~.
\label{aplopra}
\ee
$R$ introduces an additional twist that shifts $\vec{b}_1 + \vec{b}_2$ by $\vec{a}_1 \times \vec{a}_2$ in the group
composition law. As a result, the product law of three group elements is non-associative, in general.
Explicit calculation yields
\ba
& & \left(U(\vec{a}_1, ~ \vec{b}_1) U(\vec{a}_2, ~ \vec{b}_2) \right) U(\vec{a}_3, ~ \vec{b}_3) =
e^{-{i \over 2}[\vec{a}_1 \cdot (\vec{b}_2 + \vec{b}_3) - \vec{a}_2 \cdot (\vec{b}_1 - \vec{b}_3)
- \vec{a}_3 \cdot (\vec{b}_1 + \vec{b}_2) + {R \over 2} (\vec{a}_1 \times \vec{a}_2) \cdot \vec{a}_3]} \nonumber\\
& & ~~~~~~~~ \times U \left(\vec{a}_1 + \vec{a}_2 + \vec{a}_3, ~~ \vec{b}_1 + \vec{b}_2 + \vec{b}_3 -
{R \over 2} [(\vec{a}_1 \times \vec{a}_2) + (\vec{a}_1 + \vec{a}_2) \times \vec{a}_3] \right) ,
\ea
which when compared to the similar expression for $U(\vec{a}_1, \vec{b}_1) (U(\vec{a}_2, \vec{b}_2) U(\vec{a}_3, \vec{b}_3))$
it shows that
\be
\left(U(\vec{a}_1, ~ \vec{b}_1) U(\vec{a}_2, ~ \vec{b}_2) \right)
U(\vec{a}_3, ~ \vec{b}_3) = e^{-i {R \over 2} (\vec{a}_1 \times \vec{a}_2) \cdot \vec{a}_3}
U(\vec{a}_1, ~ \vec{b}_1) \left(U(\vec{a}_2, ~ \vec{b}_2) U(\vec{a}_3, ~ \vec{b}_3) \right) .
\label{noeroa}
\ee
Here, we see no trace of the 2-cocycle $\varphi_2 (\vec{a}_1 , \vec{b}_1 ; \vec{a}_2, \vec{b}_2)$ because $d \varphi_2 = 0$.

The departure from associativity obeys a certain consistency condition, which is best described in cohomological terms.
We have already examined this at the level of the Lie algebra, but now we are going to revisit it in the context of Lie group
cohomology from two different (yet complementary) points of view analogous to the preceding discussion.

First, we consider the cohomology of the Abelian group of translations in phase space with real-valued coefficients.
The value of a group cocycle depends on group elements $g_1, g_2, \cdots $, which we represent by the
canonical parameters of the group. For commutative groups, the product of two group elements is represented
by the sum of the corresponding canonical parameters. With this notation in mind, let us consider the scalar triple
product of $\vec{a}_1, ~ \vec{a}_2, ~ \vec{a}_3$ as a 3-cochain of the group of translations,
\be
\varphi_3 (\vec{a}_1 , \vec{a}_2 , \vec{a}_2)  = (\vec{a}_1 \times \vec{a}_2) \cdot \vec{a}_3 ~,
\label{tamikoe}
\ee
which appears in equation \eqn{noeroa}. For all other entries, having at least one $\vec{b}$, the value of the cochain
is taken to be zero. Although
we think of $\varphi_3$ as a cochain of the group of translations in phase space, in reality it resides in momentum
space, since it is only supported by vectors $\vec{a}$ that come multiplied with $\vec{x}$ in the group
elements \eqn{basiako} and, thus, shift the
momenta\footnote{The same cochain arises in the magnetic field analogue of the problem, which is discussed further in section 5,
but in that case it resides in configuration space, since the role of position and momenta are interchanged.}.
It can be easily checked that $\varphi_3$ is a 3-cocycle satisfying the condition
\ba
d \varphi_3 (\vec{a}_1 , \vec{a}_2 , \vec{a}_3 , \vec{a}_4) & = &
\varphi_3 (\vec{a}_2 , \vec{a}_3 , \vec{a}_4) - \varphi_3 (\vec{a}_1 + \vec{a}_2 , \vec{a}_3 , \vec{a}_4) +
\varphi_3 (\vec{a}_1 , \vec{a}_2 + \vec{a}_3 , \vec{a}_4) - \nonumber\\
& & \varphi_3 (\vec{a}_1 , \vec{a}_2 , \vec{a}_3 + \vec{a}_4) +
\varphi_3 (\vec{a}_1 , \vec{a}_2 , \vec{a}_2) = 0 ~.
\label{thatfolks}
\ea
Also, it can be verified that $\varphi_3$ is not a coboundary in real-valued cohomology of the group of
translations, since, otherwise, there would be a real-valued 2-cochain $\phi_2$ such that
$\varphi_3 (\vec{a}_1 , \vec{a}_2 , \vec{a}_3) = d \phi_2 (\vec{a}_1 , \vec{a}_2 , \vec{a}_3) =
\phi_2 (\vec{a}_2, \vec{a}_3) - \phi_2 (\vec{a}_1 + \vec{a}_2 , \vec{a}_3) + \phi_2 (\vec{a}_1 , \vec{a}_2 + \vec{a}_3) -
\phi_2 (\vec{a}_1 , \vec{a}_2)$.

The group cocycle condition of the parabolic flux models can also be derived by comparing all possible
ways four different groups elements can associate. We compute
\ba
& & \left(U(\vec{a}_1, ~ \vec{b}_1) U(\vec{a}_2, ~ \vec{b}_2) \right)
\left(U(\vec{a}_3, ~ \vec{b}_3) U(\vec{a}_4, ~ \vec{b}_4) \right) =
e^{i {R \over 4} [(\vec{a}_1 + \vec{a}_2) \cdot (\vec{a}_3 \times \vec{a}_4) -
(\vec{a}_1 \times \vec{a}_2) \cdot (\vec{a}_3 + \vec{a}_4)]} \nonumber\\
& & ~~~~ \times e^{-{i \over 2} [\vec{a}_1 \cdot (\vec{b}_2 + \vec{b}_3 + \vec{b}_4) - \vec{a}_2 \cdot
(\vec{b}_1 - \vec{b}_3 - \vec{b}_4) - \vec{a}_3 \cdot (\vec{b}_1 + \vec{b}_2 - \vec{b}_4) -
\vec{a}_4 \cdot (\vec{b}_1 + \vec{b}_2 + \vec{b}_3)]} \times \nonumber\\
& & ~~~~ U \left(\sum_{i=1}^4 \vec{a}_i , ~~ \sum_{i=1}^4 \vec{b}_i
- {R \over 2} [(\vec{a}_1 \times \vec{a}_2) + (\vec{a}_3 \times \vec{a}_4) + (\vec{a}_1 + \vec{a}_2)
\times (\vec{a}_3 + \vec{a}_4)] \right) ,
\ea
and obtain similar expressions for the other four ways of association. Then, comparing the
results, we arrive at the following general relations:
\ba
(U_1 U_2)(U_3 U_4) & = & e^{i{R \over 2} [(\vec{a}_1 + \vec{a}_2) \cdot (\vec{a}_3 \times \vec{a}_4) -
\vec{a}_3 \cdot (\vec{a}_1 \times \vec{a}_2)]} (U_1 (U_2 U_3)) U_4 ~, \nonumber\\
& = & e^{- i{R \over 2} [(\vec{a}_3 + \vec{a}_4) \cdot (\vec{a}_1 \times \vec{a}_2) -
\vec{a}_2 \cdot (\vec{a}_3 \times \vec{a}_4)]}
U_1 ((U_2 U_3) U_4) ~, \nonumber\\
& = & e^{i{R \over 2} (\vec{a}_1 + \vec{a}_2) \cdot (\vec{a}_3 \times \vec{a}_4)}
((U_1 U_2) U_3) U_4 ~, \nonumber\\
& = & e^{-i{R \over 2} (\vec{a}_3 + \vec{a}_4) \cdot (\vec{a}_1 \times \vec{a}_2)}
U_1 (U_2 (U_3 U_4)) ~,
\label{fiveways}
\ea
where, $U_1$ is used to denote $U(\vec{a}_1, ~ \vec{b}_1)$, and so on, to simplify the expressions.
Assigning each one of the five products to the corners of Mac Lane's pentagon, as depicted in
Fig.1, we observe that starting from any group element, say $(U_1 U_2)(U_3 U_4)$, and going around the pentagon
by implementing the relations \eqn{fiveways}, we arrive at the same group element without picking up a phase. This
provides a diagrammatic way to represent the five-term 3-cocycle condition \eqn{thatfolks}.

\vskip0.5cm
\begin{figure}[h]
\centering
\epsfig{file=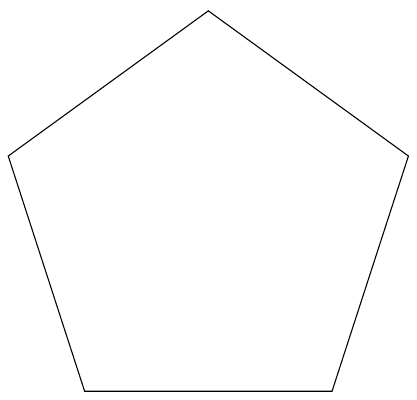,width=6cm}
\put(-120,170){$(U_1 U_2)(U_3 U_4)$ }
\put(-250,100){$U_1 (U_2 (U_3 U_4))$ }
\put(5,100){$((U_1 U_2) U_3)U_4$ }
\put(-190,-20){$U_1 ((U_2 U_3)U_4)$ }
\put(-35,-20){$(U_1 (U_2 U_3)) U_4$ }
\vskip0.7cm
\caption{\small Mac Lane's pentagon (Stasheff's associahedron $K_4$).}
\label{noE}
\end{figure}

\vskip0.5cm
\begin{figure}[h]
\centering
\vspace{-0.7cm}
\begin{minipage}[t]{.35\textwidth}
\vspace{1.32cm}
\begin{center}
\epsfig{file=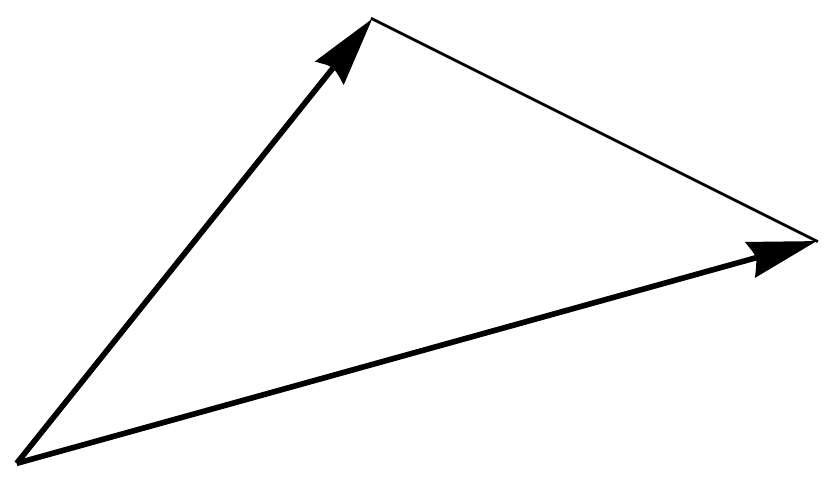, scale=.7}
\put(-80,10){$\vec{a}$}
\put(-144,50){$\vec{b}$ }
\put(-90,-90){(\hspace{.1pt}a)}
\end{center}
\end{minipage}
\hspace{1.cm}
\begin{minipage}[t]{.35\textwidth}
\begin{center}
\epsfig{file=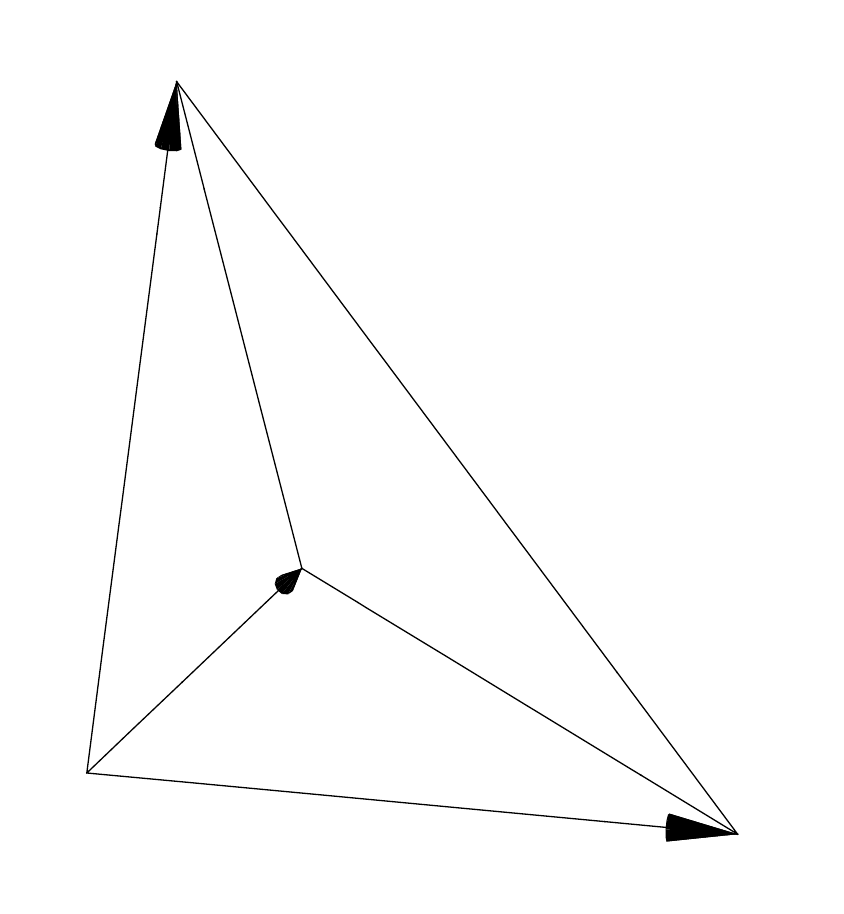, scale=0.75} 
\put(-110,11){$\vec{a}_1$ }
\put(-150,58){$\vec{a}_2$ }
\put(-170,98){$\vec{a}_3$ }
\put(-115,-27){(\hspace{.1pt}b)}
\end{center}
\end{minipage}
\vskip0.5cm
\caption{\small (a) Two vectors in phase space: the cocycle $\varphi_2 (\vec{a} , \vec{b})$ is the area of the triangle;
(b) Three vectors in momentum space: the cocycle $\varphi_3 (\vec{a}_1 , \vec{a}_2 , \vec{a}_3)$ is the volume of the tetrahedron.}
\end{figure}
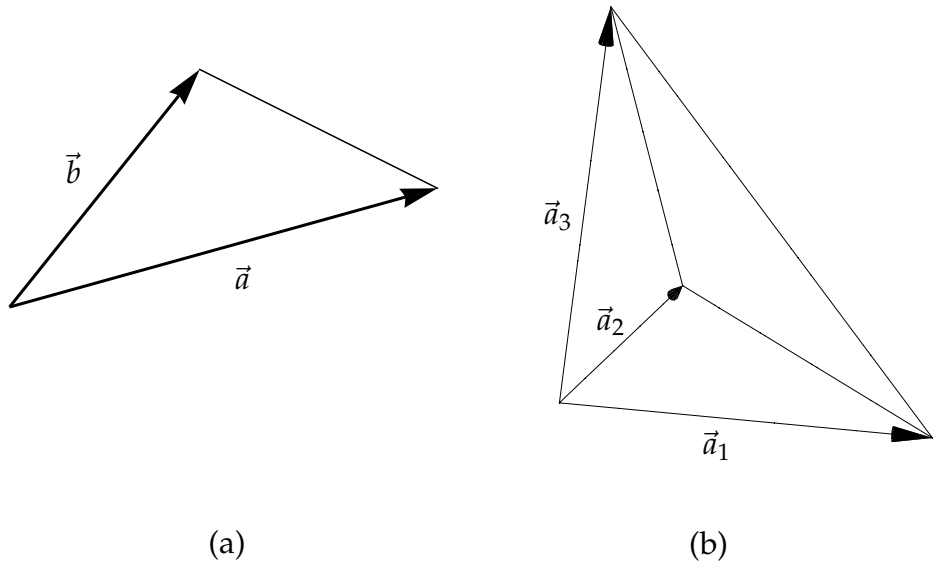

Before we proceed further a few remarks are in order on the geometric meaning of the group cocyles that were encountered
above. Fig.2a represents the triangle formed by the vectors $\vec{a}$ and $\vec{b}$ in (a two-dimensional) phase space
$(x , p)$. The 2-cocycle \eqn{aplopra} now takes the form $\varphi_2 (\vec{a} , \vec{b}) = a_1 b_2 - a_2 b_1$ and it is twice the area
of this triangle,
\be
{\rm Area} (\vec{a}, \vec{b}) = {1 \over 2} ~ |\vec{a} \times \vec{b}| ~.
\ee
Exactness of the cocycle, would simply mean that the area of the triangle could be expressed in terms of its perimeter, as
oriented sum of line elements $\phi (\vec{a})$ attached to the three sides, via
$\varphi_2 (\vec{a} , \vec{b}) = \phi (\vec{a} , \vec{b}) = \phi (\vec{a}) + \phi (\vec{b}) - \phi (\vec{a} + \vec{b})$,
but, of course, this is impossible unless the two vectors are aligned and the area vanishes. Likewise, Fig.2b represents
the tetrahedron formed by the vectors $\vec{a}_1 , ~ \vec{a}_2 , ~ \vec{a}_3$ in three-dimensional momentum space. The 3-cocycle
$\varphi_3 (\vec{a}_1 , \vec{a}_2 , \vec{a}_3) = (\vec{a}_1 \times \vec{a}_2) \cdot \vec{a}_3$ is six times the volume of this
tetrahedron,
\be
{\rm Volume} (\vec{a}_1 , \vec{a}_2 , \vec{a}_3) = {1 \over 6} ~ |(\vec{a}_1 \times \vec{a}_2) \cdot \vec{a}_3| ~.
\ee
Exactness of the cocycle would mean that the volume of the tetrahedron could be written as oriented sum of the
area of its four triangular faces, but again this is impossible unless two or all three vectors are aligned.

The obstructions carried by the group cocycles wash away when the area in phase space or the volume
in momentum space spanned by the corresponding simplices assume certain quantized values. Non-commutativity and/or
non-associativity of the corresponding group elements is restored when the phases represented by the cocycles
become integer multiples of $2 \pi$, in appropriate units. Of course, it does not mean that the cocycles trivialize, since
this occurs only for special choices of the corresponding 2- and 3-simplices and not for all.
The area quantum assigned to $\varphi_2$ is nothing else but Planck's cell in phase space
(equal to $2 \pi$ in units of $\hbar$), which accounts for the fuzziness of points and measures the "size" of
non-commutativity in phase space (and subsequently of the Moyal star-product that will be discussed later).
Any two triangles differing by an integer multiple of Planck's cell experience the same "amount" of non-commutativity
among the group elements that span it.
Likewise, the quantum cells in momentum space assigned to $\varphi_3$ "measure" the fuzziness of points due to
non-associativity in the parabolic flux models (which is reflected to the non-associativity of the corresponding
star-product that will also be discussed later). These remarks provide physical meaning to the cocycles, and their
corresponding simplices, and they are in agreement with the uncertainty relations for
non-commutativity/non-associativity derived some time ago in reference \cite{Lust:2010iy}.

There is yet another useful way to characterize the non-associativity of parabolic flux models, using the group
cohomology of the Heisenberg-Weyl group $G_{W}$ with cochains taking values in the Heisenberg algebra ${\bf g}$. It is the
group theory analog of Chevalley-Eilenberg cohomology for the Heisenberg algebra that was discussed earlier.

Recall that the Heisenberg algebra ${\bf g}$ carries
a representation of $G_W$, which is inherited from the adjoint action of the algebra and it is realized by
conjugation, as $\pi (g^{\prime}) {\bf g} = g^{\prime -1} {\bf g} g^{\prime}$ for all $g^{\prime} \in G_W$. In this
context, it is also useful to think of the group composition law \eqn{groupcompos} as
\be
U_W (g_1) U_W (g_2) = e^{-i {R \over 2} \varphi_2 (g_1 , g_2)} U_W (g_1 g_2) ~,
\label{tsitare}
\ee
where $U_W (g)$ extends the action of the group of translations $U$ to the Heisenberg-Weyl group in the
obvious way, including the central element $\mathbb{1}$ among the generators,
\be
U_W (g) = e^{i (\vec{a} \cdot \vec{x} + \vec{b} \cdot \vec{p} + c \mathbb{1})} ~.
\ee
The "phase" appearing in \eqn{tsitare} does not assume real values, as in ordinary quantum mechanics,
but it is a 2-cochain in the group cohomology
of $G_W$ with coefficients in the Lie algebra ${\bf g}$. Then, equation \eqn{tsitare} is simply a reformulation of
the original composition law \eqn{groupcompos} provided that the 2-cochain is chosen as
\be
\varphi_2 (g_1 , g_2) = (\vec{a}_1 \times \vec{a}_2) \cdot \vec{p} ~.
\label{cocha135}
\ee
The normalization factor has already been extracted and it appears in the exponent in equation \eqn{tsitare}, but it
is irrelevant for the purposes of the present discussion.

Computing the action of the coboundary operator $d$ on $\varphi_2 \in C^2 (G_W , {\bf g})$, we find
\be
d \varphi_2 (g_1 , g_2 , g_3) = \pi (g_1) \varphi_2 (g_2 , g_3) - \varphi_2 (g_1 g_2 , g_3) + \varphi_2 (g_1 , g_2 g_3)
- \varphi_2 (g_1 , g_2) ~.
\label{trirona}
\ee
Using the Baker-Campbell-Hausdorff formula for the Heisenberg algebra, we obtain
\be
\pi (g_1) \varphi_2 (g_2 , g_3) = \varphi_2 (g_2 , g_3) -i [\vec{a}_1 \cdot \vec{x} + \vec{b}_1 \cdot \vec{p} , ~~
\varphi_2 (g_2, g_3)] = \varphi_2 (g_2 , g_3) + \vec{a}_1 \cdot (\vec{a}_2 \times \vec{a}_3)
\ee
and, thus, the final result reads
\be
(\vec{a}_1 \times \vec{a}_2) \cdot \vec{a}_3  = d \varphi_2 (g_1 , g_2 , g_3) ~,
\ee
expressing the obstruction to associativity as a coboundary in the Lie algebra valued group cohomology
of $G_W$. All other terms emerging from equation \eqn{trirona} cancel against each other.

Finally, we mention briefly that all other faces of the toroidal flux model can be treated similarly
by appropriate relabeling of the coordinates and momenta, placing tildes wherever is necessary, as in the
Lie algebra cohomology. Passing directly to the group cohomology in double phase space, we note that
the group elements \eqn{basiako} generalize to
\be
U (\vec{a}, ~ \vec{b} ; \vec{c} , ~ \vec{d}) = e^{i(\vec{a} \cdot \vec{x} + \vec{b} \cdot \vec{p} +
\vec{c} \cdot \vec{\tilde{x}} + \vec{d} \cdot \vec{\tilde{p}})}
\ee
and their composition law involves a 2-cochain that depends linearly on the momenta or its dual, depending
on the chosen duality frame. Thus, in the cohomology of the Lie group $G_W \times \tilde{G}_W$ with values in the
Lie algebra ${\bf g} \oplus {\bf \tilde{g}}$, the cochain assumes the following form in the $H, ~ Q$ and $R$-frames
\ba
H-{\rm flux}: ~~~~ & & \varphi_2 (g_1, g_2 ; \tilde{g}_1, \tilde{g}_2) = (\vec{c}_1 \times \vec{c}_2) \cdot \vec{\tilde{p}} ~,
\label{yamawa1} \\
Q-{\rm flux}: ~~~~ & & \varphi_2 (g_1, g_2 ; \tilde{g}_1, \tilde{g}_2) = (\vec{a}_1 \times \vec{a}_2) \cdot \vec{\tilde{p}} ~,
\label{yamawa2} \\
R-{\rm flux}: ~~~~ & & \varphi_2 (g_1, g_2 ; \tilde{g}_1, \tilde{g}_2) = (\vec{a}_1 \times \vec{a}_2) \cdot \vec{p} ~.
\label{yamawa3}
\ea
whereas in the $f$-frame that involves non-trivial commutation relations among the coordinates and their dual counterparts
the result is
\be
f-{\rm flux}: ~~~~~~~ \varphi_2 (g_1, g_2 ; \tilde{g}_1, \tilde{g}_2) = [(\vec{a}_1 \times \vec{c}_2) -
(\vec{a}_2 \times \vec{c}_1)] \cdot \vec{\tilde{p}} ~.
\label{yamawa4}
\ee

These results will be particularly useful in the next section aiming at the systematic construction of
star-products in double phase space that are valid in all duality frames. Naturally, we expect that the
star-product of functions in double phase will be non-associative and depend on $\vec{p}$ as well as
$\vec{\tilde{p}}$.

\section{Star-products and double field theory phase space}
\setcounter{equation}{0}

This section is devoted to the derivation of the star-product of functions in phase space. First, we
overview the construction of the Moyal product, which is familiar from elementary quantum mechanics,
and, then, we generalize it to the parabolic string model using the $R$-flux frame. The results are
extended to all other frames and finally unified in a duality invariant way using the double field
theory phase space.

\subsection{Moyal product and 2-cocycles}

We recall the construction of the star-product (also known as Moyal product)
among functions on the phase space $\mathbb{R}^{2n}$ with position and momentum coordinates $(x^i , p^i)$, letting
$i = 1, 2, \cdots , n$, in general, without extra effort. This product, which is denoted by $(f_1 \star f_2)(x, p)$,
provides a non-commutative but associative composition law that is isomorphic to the product of operators
$\hat{F}_1 \cdot \hat{F}_2$, which represent the corresponding phase space functions $f_1$ and $f_2$
in quantum mechanics with a given factor ordering prescription. The presentation is self-contained
emphasizing the cohomological aspects of the problem that need to be generalized later to encompass
the phase space structure of non-geometric string backgrounds.

The operators representing position and momenta obey the Heisenberg commutation
relations
\be
[\hat{x}^i , ~ \hat{p}^j] = i \delta^{ij} ~.
\ee
As such they provide a central extension of the
Abelian algebra of translations in phase space. Denoting by $T_I$ with $I= 1, 2, \cdots , 2n$ the generators
of translations in $\mathbb{R}^{2n}$, this means that there is a real-valued anti-symmetric bilinear function
of the generators $c_2(T_I , T_J)$ satisfying the 2-cocycle condition for all generators,
\be
c_2([T_I , ~ T_J], ~ T_K) + c_2([T_J , ~ T_K], ~ T_I) + c_2([T_K , ~ T_I], ~ T_J) = 0,
\ee
so that the Heisenberg algebra is described as $[T_I , ~ T_J] = i c_2(T_I, T_J)$ for appropriate choice
of $c_2(T_I, T_J)$. If the cochain were exact, having the special form $c_2(T_I, T_J)= \theta ([T_I, T_J])$,
it would be identically zero by the Abelian nature of the algebra, $[T_I , T_J] =0$. For non-Abelian
algebras exact cochains are not necessarily zero, but they can be removed by shifting all generators
$T_I$ by the constant elements $\theta (T_I)$. In the present case, a non-trivial cocycle arises for
the choice $c_2(T_{i}, T_{n+i}) = 1$, up to multiplication, while it is zero for all other pairs of
algebra generators. Setting $T_i = \hat{x}^i$ and $T_{n+i} = \hat{p}^i$ for all
$i = 1, 2, \cdots , n$ we obtain the standard description of the Heisenberg algebra as central extension
of the algebra of translations.

Next, we exponentiate the action of the Lie algebra by considering the group elements with canonical
parameters $\vec{a}$ and $\vec{b}$,
\be
\hat{U}(\vec{a}, ~ \vec{b}) = e^{i(\vec{a} \cdot \hat{\vec{x}} + \vec{b} \cdot \hat{\vec{p}})} ~,
\ee
which satisfy the following product relation, via the Baker-Campbell-Hausdorff formula,
\be
\hat{U}(\vec{a}_1, ~ \vec{b}_1) \hat{U}(\vec{a}_2, ~ \vec{b}_2) =
e^{-{i \over 2} (\vec{a}_1 \cdot \vec{b}_2 - \vec{a}_2 \cdot \vec{b}_1)}
\hat{U} (\vec{a}_1 + \vec{a}_2, ~ \vec{b}_1 + \vec{b}_2 ) ~ .
\label{projrep}
\ee
The phase $\varphi_2 (\vec{a}_1, \vec{b_1} ; \vec{a}_2, \vec{b}_2) =
\vec{a}_1 \cdot \vec{b}_2 - \vec{a}_2 \cdot \vec{b}_1$ is the real-valued 2-cocycle of the
Abelian group of translations in phase space that was discussed before, i.e., $\hat{U} (\vec{a} , \vec{b})$ is only
a projective representation of the group of translations satisfying the associativity
condition
\be
\left(\hat{U} (\vec{a}_1 , \vec{b}_1) \hat{U} (\vec{a}_2 , \vec{b}_2) \right) \hat{U} (\vec{a}_3 , \vec{b}_3) =
\hat{U} (\vec{a}_1 , \vec{b}_1) \left(\hat{U} (\vec{a}_2 , \vec{b}_2) \hat{U} (\vec{a}_3 , \vec{b}_3) \right) ~.
\ee
If the group cochain were exact, it would be rewritten in terms of a single function $\phi (\vec{a}, \vec{b})$,
as $\varphi (\vec{a}_1, \vec{b_1} ; \vec{a}_2, \vec{b}_2) = \phi (\vec{a}_1, \vec{b}_1) + \phi (\vec{a}_2, \vec{b}_2)
- \phi (\vec{a}_1 + \vec{a}_2 , \vec{b}_1 + \vec{b}_2)$,
and, hence, it could be absorbed in the phase of $\hat{U} (\vec{a}, \vec{b})$ systematically, leading to an
ordinary representation of the group of translations. Of course, this is not the case here, as we are dealing with a
genuine projective representation of the group set by Planck's constant $\hbar$ that is normalized to $1$.

Let us now consider the space of all (suitably continuous) functions on the classical phase space $\mathbb{R}^{2n}$.
It is convenient to decompose any such function $f(\vec{x}, \vec{p})$ in modes as follows,
\be
f(\vec{x}, \vec{p}) = { 1 \over (2 \pi)^n} \int d^n a d^n b ~ \tilde{f} (\vec{a}, \vec{b})
e^{i(\vec{a} \cdot \vec{x} + \vec{b} \cdot \vec{p})} ~,
\ee
where $\tilde{f} (\vec{a}, \vec{b})$ is the Fourier transform of $f(\vec{x}, \vec{p})$,
\be
\tilde{f}(\vec{a}, \vec{b}) = { 1 \over (2 \pi)^n} \int d^n x d^n p ~ f (\vec{x}, \vec{p})
e^{- i(\vec{a} \cdot \vec{x} + \vec{b} \cdot \vec{p})} ~.
\ee
Then, Weyl's correspondence rule assigns a Hermitean operator $\hat{F}$ to any given function $f$, as
\be
\hat{F}(\hat{\vec{x}}, \hat{\vec{p}}) = { 1 \over (2 \pi)^n} \int d^n a d^n b ~ \tilde{f} (\vec{a}, \vec{b})
\hat{U} (\vec{a} , \vec{b}) ~,
\ee
using $\hat{U} (\vec{a} , \vec{b})$ to represent $e^{- i(\vec{a} \cdot \vec{x} + \vec{b} \cdot \vec{p})}$
upon quantization. The correspondence is one-to-one, thus taking care of the factor ordering ambiguities
that arise, in general, for arbitrary functions on the phase space. Other correspondence rules can also be
used at will, but the main construction will remain essentially the same. In any case,
the resulting star-product algebra is unique, since different factor ordering prescriptions constitute
a change of base and, thus, the resulting algebraic structures are isomorphic.

The product of any two operators $\hat{F}_1$ and $\hat{F}_2$ assumes the following representation in terms
of Weyl's correspondence rule:
\be
\hat{F}_1 \cdot \hat{F}_2 = { 1 \over (2 \pi)^{2n}} \int d^n a_1 d^n b_1 d^n a_2 d^n b_2 ~ \tilde{f}_1 (\vec{a}_1, \vec{b}_1)
\tilde{f}_2 (\vec{a}_2, \vec{b}_2) \hat{U} (\vec{a}_1, ~ \vec{b}_1) \hat{U} (\vec{a}_2, ~ \vec{b}_2) ~.
\ee
Using the projective representation \eqn{projrep} of the translation group, the result takes the form
\be
\hat{F}_1 \cdot \hat{F}_2 = { 1 \over (2 \pi)^{2n}} \int d^n a_1 d^n b_1 d^n a_2 d^n b_2 ~ \tilde{f}_1 (\vec{a}_1, \vec{b}_1)
\tilde{f}_2 (\vec{a}_2, \vec{b}_2) e^{-{i \over 2} (\vec{a}_1 \cdot \vec{b}_2 - \vec{a}_2 \cdot \vec{b}_1)}
\hat{U} (\vec{a}_1 + \vec{a}_2, ~ \vec{b}_1 + \vec{b}_2 ) ~
\ee
and, thus, by employing once more Weyl's correspondence rule, the corresponding function in phase space is
\be
f_1 \star f_2 = { 1 \over (2 \pi)^{2n}} \int d^n a_1 d^n b_1 d^n a_2 d^n b_2 ~ \tilde{f}_1 (\vec{a}_1, \vec{b}_1)
\tilde{f}_2 (\vec{a}_2, \vec{b}_2) e^{-{i \over 2} (\vec{a}_1 \cdot \vec{b}_2 - \vec{a}_2 \cdot \vec{b}_1)}
e^{i [(\vec{a}_1 + \vec{a}_2) \cdot \vec{x} +  (\vec{b}_1 + \vec{b}_2) \cdot \vec{p}]} ~.
\ee
This is the defining relation of the Moyal star-product among any two functions on
phase space.

The star-product can be rewritten via Fourier transform as follows, setting for convenience
$\vec{a}_1 + \vec{a}_2 = \vec{a}$, $\vec{b}_1 + \vec{b}_2 = \vec{b}$ and $\vec{a}_2 = \vec{a}^{\prime}$,
$\vec{b}_2 = \vec{b}^{\prime}$,
\be
(f_1 \star f_2) (\vec{x}, \vec{p}) = { 1 \over (2 \pi)^n} \int d^n a d^n b ~ (\tilde{f}_1 \odot
\tilde{f}_2) (\vec{a}, \vec{b}) e^{i(\vec{a} \cdot \vec{x} + \vec{b} \cdot \vec{p})} ,
\ee
where
\be
(\tilde{f}_1 \odot \tilde{f}_2) (\vec{a} , \vec{b}) = { 1 \over (2 \pi)^n}
\int d^n a^{\prime} d^n b^{\prime} ~ \tilde{f}_1 (\vec{a} - {\vec{a}}^{\prime},
\vec{b} - {\vec{b}}^{\prime}) \tilde{f}_2 ({\vec{a}}^{\prime} , {\vec{b}}^{\prime})
e^{-{i \over 2} (\vec{a} \cdot \vec{b}^{\prime} - \vec{a}^{\prime} \cdot \vec{b})}
\label{twcovds}
\ee
is the convolution among $\tilde{f}_1$ and $\tilde{f}_2$ twisted by the 2-cocycle of the
Abelian group of translations in phase space. Conversely, one also has
\be
(\tilde{f}_1 \odot \tilde{f}_2) (\vec{a}, \vec{b}) = { 1 \over (2 \pi)^n} \int d^n x d^n p ~
(f_1 \star f_2) (\vec{x}, \vec{p}) e^{- i(\vec{a} \cdot \vec{x} + \vec{b} \cdot \vec{p})} ~.
\ee
The Moyal star-product is non-commutative but associative thanks to the 2-cocycle condition. If the
2-cochain were exact, the star product would be commutative, but, of course, this is not the case when $\hbar \neq 0$.

Completing the presentation, we expand the twisted convolution around the standard one and perform the
integrations in all $\vec{a}$- and $\vec{b}$-variables. This leads to the appearance of derivative terms with respect
to the phase space coordinates $\vec{x}$ and $\vec{p}$ that correct the usual product of functions and turns it into
non-commutative. The end result is neatly summarized as follows,
\be
(f_1 \star f_2) (\vec{x}, \vec{p}) = e^{{i \over 2} \left(\vec{\nabla}_{x_1} \cdot \vec{\nabla}_{p_2} -
\vec{\nabla}_{x_2} \cdot \vec{\nabla}_{p_1} \right)} f_1 (\vec{x}_1, \vec{p}_1)
f_2 (\vec{x}_2, \vec{p}_2) |_{\vec{x}_1 = \vec{x}_2 = \vec{x}; ~ \vec{p}_1 = \vec{p}_2 = \vec{p}} ~,
\label{moyalia3}
\ee
where $\vec{x}$ and $\vec{p}$ are implicitly assumed to be $n$-dimensional. Then, we obtain
\be
(f_1 \star f_2) (\vec{x}, \vec{p}) = (f_1 \cdot f_2)(\vec{x}, \vec{p}) + {i \over 2} \{f_1 , ~ f_2 \} + \cdots ~.
\label{expans}
\ee
The first correction term is the Poisson bracket among the phase space functions $f_1$ and $f_2$,
whereas the dots denote all higher derivative terms in $\vec{x}$ and $\vec{p}$ that arise from the power series expansion
of the exponential.

Reinstating Planck's constant, which so far was normalized to $1$, we obtain $[\hat{x}^i, ~ \hat{p}^j] =
i \hbar \delta^{ij}$ and the group cocycle $\varphi_2 (\vec{a}_1, \vec{b_1} ; \vec{a}_2, \vec{b}_2) =
\vec{a}_1 \cdot \vec{b}_2 - \vec{a}_2 \cdot \vec{b}_1$ acquires a factor of $\hbar$. Then, the derivative expansion
of the star-product can be organized as power series in Planck's constant,
\be
(f_1 \star f_2) (\vec{x}, \vec{p}) = (f_1 \cdot f_2)(\vec{x}, \vec{p}) + \hbar K_1 (f_1, f_2) + {\hbar}^2 K_2 (f_1, f_2) + \cdots ~,
\ee
with $K_0 (f_1, f_2) = f_1 \cdot f_2$ and $K_1 (f_1 , f_2) = (i/2) \{f_1 , f_2 \}$, as before. $K_n (f_1 , f_2)$
are bilinear terms containing $n$ derivatives with respect to $\vec{x}$ and $n$ derivatives with respect to $\vec{p}$ ($2n$
number of derivatives in total). As consequence of associativity, the following relations are valid order by order
in powers of $\hbar$ for all values $r = 0, 1, 2, \cdots $,
\be
\sum_{n + m = r} K_n (K_m (f_1 , f_2) , f_3) - K_n(f_1, K_m (f_2, f_3)) = 0 ~.
\ee
They can be interpreted as 2-cocycle conditions for having an extension of the commutative algebra of functions
on the phase space by the algebra itself; in this case we have a deformation of an associative algebra and the corresponding
cocycle is properly described in terms of Hochschild cohomology in the spirit of Gerstenhaber \cite{gerstenhaber}.

A closely related Lie algebraic structure is introduced by the Moyal bracket on the phase space functions, as
\be
\{\{f_1 , ~ f_2 \}\} = -{i \over \hbar} (f_1 \star f_2 - f_2 \star f_1) ~,
\ee
which provides a deformation of the Poisson bracket algebra by higher derivative terms in $\vec{x}$ and $\vec{p}$.
By construction, it is isomorphic to the algebra of Hermitean operators under the commutator
$-(i/\hbar) [\hat{F}_1, ~ \hat{F}_2]$. Associativity of the Moyal star-product implies immediately that
\be
\{\{f_1 , ~ f_2 \star f_3 \}\} = f_2 \star \{\{f_1 , ~ f_3 \}\} + \{\{f_1 , ~ f_2 \}\} \star f_3 ~.
\ee
Then, $\mathbb{X}_f = \{\{ f, ~ \cdot ~ \}\}$ acts as a derivation on the Moyal product algebra, satisfying
Leibnitz rule, and, as such, it provides the quantum analogue (with higher order differential operators) of a Hamiltonian
vector field associated to the function $f$, modulo constants, which is tangent to non-commutative phase space. One has
the following relation among such vector fields, $[\mathbb{X}_{f_1} , ~ \mathbb{X}_{f_2}] = \mathbb{X}_{\{\{f_1 , f_2 \}\}}$.
This should be compared to the ordinary Hamiltonian formalism on commutative phase space, where the Hamiltonian
vector fields $X_f = \{f , ~ \cdot ~ \}$ generated by the Poisson bracket satisfy a similar relation
$[X_{f_1} , ~ X_{f_2}] = X_{\{f_1 , f_2 \}}$. The first order operators $X_f$ act like ordinary vector fields in
commutative geometry, and they form the algebra of volume preserving diffeomorphisms in phase space. The
Moyal algebra generated by $\mathbb{X}_f$ provides a consistent quantization of all these notions of classical
symplectic geometry.

The non-commutativity of the star-product arising from the derivative terms in the expansion \eqn{expans}
is intimately related to the fuzziness of quantum phase space. Indeed, if two functions vanish at a point their
ordinary product will also vanish at that point, but their star-product will not be zero; the star-product
will vanish if and only if all derivatives of the two functions vanish at that point, in which case the functions
will be zero everywhere. This provides a rather precise way to think of non-commutative geometry in terms of
non-commutative algebraic structures on the space of functions. It should be noted that
the same method applies to the algebraic description of other non-commutative spaces that do not necessarily
have the interpretation of phase space. In those cases one should simply think of the momenta $p^i$ as some
additional non-commuting coordinates, as in the example of a non-commutative plane with coordinates $x^1$ and $x^2$
satisfying the relations $[x^i , ~ x^j] = i \epsilon^{ij}$. Also, (some of) the coordinates can be periodic, as
in the case of a phase space with the topology of cylinder or that of a non-commutative space with the topology of
torus; one simply has to consider periodic functions along those circles, and not use the angular coordinates
themselves, in order to extend the validity of the formalism -- recall that there is no Hermitean operator that
can be assigned to an angle.

Next, we extend the scope of this algebraic description to non-associative geometry as well, based on suitably
defined star-products. As will be seen later, the Moyal product is not just a toy problem to motivate more
general constructions, but it also becomes integral part of the non-associative star-product
in the double phase space of the parabolic flux models.

\subsection{Non-associative star-product and 3-cocycles}

We turn to the commutation relations for the coordinates and momenta of the parabolic flux models, which
are conveniently written in smeared form, using the three-vectors $\vec{a}$ and $\vec{b}$ to smear
$\vec{x}$ and $\vec{p}$, respectively,
\be
[\vec{a}_1 \cdot \vec{x} , ~~ \vec{a}_2 \cdot \vec{x}] = i R (\vec{a}_1 \times \vec{a}_2) \cdot \vec{p} ~,
~~~~~~ [\vec{a} \cdot \vec{x} , ~~ \vec{b} \cdot \vec{p}] = i ~ \vec{a} \cdot \vec{b} ~,
\ee
whereas the associator assumes the smeared form
\be
[\vec{a}_1 \cdot \vec{x} , ~~ \vec{a}_2 \cdot \vec{x} , ~~ \vec{a}_3 \cdot \vec{x}] = 3i R
(\vec{a}_1 \times \vec{a}_2) \cdot \vec{a}_3
\ee
and it involves the scalar triple product of the vectors $\vec{a}_1$, $\vec{a}_2$ and $\vec{a}_3$. The obstruction has the
form $(\vec{a}_1 \times \vec{a}_2) \cdot \vec{a}_3$ and it is immediately recognized to be the 3-cocycle of the Abelian
group of translations in phase space.

Here, it appears as if we are dealing only with the $R$-flux model, but actually one can extend the formalism to
any dual face using the variables $\vec{x}$ to denote collectively the coordinates or the dual coordinates of the 3-torus,
depending on the chosen duality frame, and, likewise, $\vec{p}$ to denote collectively the momenta or the dual momenta
along the toroidal directions. Table 1 helps again to keep track of the models and the labels.
We will do this properly in the next subsection, using the double phase space description, in order
to produce a duality invariant framework for the parabolic flux models.

We are going to introduce a star-product among the functions of the phase space $(x^i , p^i)$ that resembles the
Moyal product for the Heisenberg algebra. As first step, we exponentiate the action of the position and momentum
generators and consider the group elements
\be
\hat{U}(\vec{a}, ~ \vec{b}) = e^{i(\vec{a} \cdot \hat{\vec{x}} + \vec{b} \cdot \hat{\vec{p}})} ~.
\ee
They are the same group elements \eqn{basiako} that were introduced earlier for the $R$-flux model. Putting hats does not
mean that we are representing these elements as operators acting on Hilbert space, but it is a useful booking notation to
differentiate them from the classical functions $e^{i(\vec{a} \cdot \vec{x} + \vec{b} \cdot \vec{p})}$ that will be used
in the following.

As in the previous subsection, we consider all (suitably continuous) functions
$f(\vec{x}, \vec{p})$ on six-dimensional phase space and decompose them as
\be
f(\vec{x}, \vec{p}) = { 1 \over (2 \pi)^3} \int d^3 a d^3 b ~ \tilde{f} (\vec{a}, \vec{b})
e^{i(\vec{a} \cdot \vec{x} + \vec{b} \cdot \vec{p})} ~,
\ee
using the Fourier transformed functions $\tilde{f} (\vec{a}, \vec{b})$.
We are going to make formal use of Weyl's correspondence rule by assigning $\hat{F}$ to any given function $f$, as follows,
\be
\hat{F} = { 1 \over (2 \pi)^3} \int d^3 a d^3 b ~ \tilde{f} (\vec{a}, \vec{b}) \hat{U} (\vec{a} , \vec{b}) ~,
\ee
hereby replacing $e^{i(\vec{a} \cdot \vec{x} + \vec{b} \cdot \vec{p})}$ by $\hat{U} (\vec{a} , \vec{b})$.
Here, again, $\hat{F}$ is not meant to be an operator acting on Hilbert space, since there is no way to realize the commutation
relations among the coordinates and momenta by the rules of quantum mechanics. $\hat{F}$ is only an auxiliary quantity
that arises by smearing $\hat{U} (\vec{a} , \vec{b})$ with $\tilde{f} (\vec{a}, \vec{b})$ and, as such, it exists
at the same level of rigor as $U$ itself.

Multiplying any two such objects, $\hat{F}_1$ and $\hat{F}_2$, we obtain the following product form,
\be
\hat{F}_1 \cdot \hat{F}_2 = { 1 \over (2 \pi)^{6}} \int d^3 a_1 d^3 b_1 d^3 a_2 d^3 b_2 ~ \tilde{f}_1 (\vec{a}_1, \vec{b}_1)
\tilde{f}_2 (\vec{a}_2, \vec{b}_2) \hat{U} (\vec{a}_1, ~ \vec{b}_1) \hat{U} (\vec{a}_2, ~ \vec{b}_2) ~,
\ee
which can be manipulated, using the relation \eqn{groupcompos}, and be brought in a form that is more
convenient to work with,
\ba
\hat{F}_1 \cdot \hat{F}_2 & = & { 1 \over (2 \pi)^{6}} \int d^3 a_1 d^3 b_1 d^3 a_2 d^3 b_2 ~ \tilde{f}_1 (\vec{a}_1, \vec{b}_1)
\tilde{f}_2 (\vec{a}_2, \vec{b}_2) e^{-{i \over 2} (\vec{a}_1 \cdot \vec{b}_2 - \vec{a}_2 \cdot \vec{b}_1)} \nonumber\\
& & ~~~~~~~~~~~~~~ \times \hat{U} \left(\vec{a}_1 + \vec{a}_2, ~ \vec{b}_1 + \vec{b}_2 -  {R \over 2}
(\vec{a}_1 \times \vec{a}_2) \right) .
\ea
Making formal use of Weyl's correspondence rule once more, we arrive at the following star-product among any two functions
$f_1$ and $f_2$ in phase space, which is written in integral form as
\ba
(f_1 \star_p f_2) (\vec{x} , \vec{p}) & = & { 1 \over (2 \pi)^6} \int d^3 a_1 d^3 b_1 d^3 a_2 d^3 b_2 ~ \tilde{f}_1 (\vec{a}_1, \vec{b}_1)
\tilde{f}_2 (\vec{a}_2, \vec{b}_2) e^{-{i \over 2} (\vec{a}_1 \cdot \vec{b}_2 - \vec{a}_2 \cdot \vec{b}_1)} \nonumber\\
& & ~~~~~~~~~~~~~~ \times e^{-i{R \over 2} (\vec{a}_1 \times \vec{a}_2) \cdot \vec{p}}
~ e^{i [(\vec{a}_1 + \vec{a}_2) \cdot \vec{x} +  (\vec{b}_1 + \vec{b}_2) \cdot \vec{p}]} ~.
\label{newstarprodu}
\ea

All steps taken above are formal, but they are analogous to those taken in quantum mechanics,
thus providing complete justification for the definition \eqn{newstarprodu}.
Compared to the Moyal product, there is an additional twist, due to the appearance of the factor
${\rm exp} (-i R (\vec{a}_1 \times \vec{a}_2) \cdot \vec{p}/2)$ in the integrand, which introduces momentum
dependence in the star-product of any two functions of $\vec{x}$ (and, of course, additional momentum dependence
in the product of any two phase space functions). This is indicated by the subscript in $\star_p$
for distinction. The extra factor is the 2-cochain \eqn{cocha135} that was discussed earlier in the context of
group cohomology and it is responsible for the non-associativity of the newly defined star-product.

Performing the necessary integrations over the $\vec{a}$- and $\vec{b}$-variables, we arrive at the following expression
for the star-product,
\be
(f_1 \star_p f_2) (\vec{x}, \vec{p}) = e^{i {R \over 2} ~ \vec{p} \cdot (\vec{\nabla}_{x_1} \times \vec{\nabla}_{x_2})}
e^{{i \over 2} \left(\vec{\nabla}_{x_1} \cdot \vec{\nabla}_{p_2} -
\vec{\nabla}_{x_2} \cdot \vec{\nabla}_{p_1} \right)} f_1 (\vec{x}_1, \vec{p}_1)
f_2 (\vec{x}_2, \vec{p}_2) |_{\vec{x}; ~ \vec{p}} ~,
\label{moyalia13}
\ee
which is analogous to formula \eqn{moyalia3} for the Moyal product. The restriction means that one sets
$\vec{x}_1 = \vec{x}_2 = \vec{x}$ and $\vec{p}_1 = \vec{p}_2 = \vec{p}$ after computing the derivatives with respect
to the arguments of the two functions. The Fourier transform of $f_1 \star_p f_2$ is a twisted convolution
of $\tilde{f}_1$ and $\tilde{f}_2$, similar to \eqn{twcovds}, but it is now further deformed with the
$p$-dependent 2-cochain \eqn{cocha135}.

A more compact way to express the result is provided in terms of the matrix
\begin{equation}
\theta^{IJ} (p) = \begin{pmatrix} R^{ijk}p_k & & \delta^i_j \\
  &   &  \\
-\delta^j_i &  & 0   \end{pmatrix} ; ~~~~~~~~ R^{ijk} = R ~ \epsilon^{ijk} ~,
\label{twistabrac}
\end{equation}
whose elements depend on the momenta. The indices take values $I,J=1,\dots ,6$. $\theta^{IJ} (p)$ introduces
a twisted Poisson structure in phase space. Its blocks are labeled by $x$ and $p$ and each one of them is a
$3 \times 3$ matrix. We write
\begin{equation}
(f_1 \star_p f_2) (\vec{x}, \vec{p}) = e^{{i \over 2} \theta^{IJ}(p)\,
\partial_{I} \otimes \partial_J} \, (f_1 \otimes f_2) |_{\vec{x}; ~ \vec{p}} \, .
\label{tsajio}
\end{equation}

The same expression \eqn{tsajio} together with the matrix \eqn{twistabrac} appeared in the literature recently, \cite{Mylonas:2012pg},
while considering the quantization of a membrane $\sigma$-model via Kontsevich's deformation approach \cite{kontsev} together
with the cocycle \eqn{tamikoe}.
Here, we derived that formula, first in integral form, in a direct way based on group multiplication and group cohomology 
following Weyl's correspondence rule, thus providing complete justification for the
definition of the $\star_p$-product given by those authors.
One is tempted to think of it as first quantization of closed strings in non-geometric backgrounds, as the authors of
reference \cite{Mylonas:2012pg} do, but, as a matter of terminology, one should use these words with caution. 
Namely, we prefer to think of the star-product
as {\em substitute} for canonical quantization, since there is no equivalent formulation of the problem in terms of operators acting
on Hilbert space in this case, and reserve the term quantization only for representations of associative algebras. 
In any case, our presentation complements nicely their work.
The $\star_p$-product is defined for functions on the entire phase space,
but it can also be restricted to functions of $x$ and still
yield a non-commutative non-associative structure. It should be contrasted to the properties of the ordinary Moyal product of
functions in phase space that do not provide that last option.

There is a $\star_p$-bracket defined as $\{\{f_1 , ~ f_2 \}\}_p = -i (f_1 \star_p f_2 - f_2 \star_p f_1)$ for all functions
in phase space that reproduces the twisted Poisson bracket relations among the coordinates and momenta,
\be
\{\{x^i , ~ x^j\}\}_p = R^{ijk} p^k ~, ~~~~~~ \{\{x^i , ~ p^j\}\}_p = \delta^{ij} ~,
\ee
as expected on general grounds. It is the analogue of Moyal bracket and, as such, it can be regarded as bona fide deformation of
the twisted Poisson structure in classical phase space. Although these algebraic structures arise in the parabolic flux model,
they can be easily extended to problems with different topologies by relaxing the periodicity conditions on the
coordinates. It is rather intriguing that such structures can also be used to
describe the dynamics of point-particles in $\mathbb{R}^3$ in the background of magnetic charges.

The $\star_p$-product of any two functions of $x$ is a function that depends on $x$ as well as $p$. The product of any three
functions of $x$ turns out to be independent of $p$, but the outcome depends on the way that the three functions associate.
A simple calculation shows that
\be
\left((f_1 \star_p f_2) \star_p f_3\right) (\vec{x}) = e^{{R \over 4} (\vec{\nabla}_{x_1} \times \vec{\nabla}_{x_2})
\cdot \vec{\nabla}_{x_3}} f_1 (\vec{x}_1) f_2 (\vec{x}_2) f_3 (\vec{x}_3) |_{\vec{x}_1 = \vec{x}_2 = \vec{x}_3 = \vec{x}} ~,
\label{tsinghi}
\ee
whereas
\be
\left(f_1 \star_p (f_2 \star_p f_3) \right) (\vec{x}) = e^{- {R \over 4} (\vec{\nabla}_{x_1} \times \vec{\nabla}_{x_2})
\cdot \vec{\nabla}_{x_3}} f_1 (\vec{x}_1) f_2 (\vec{x}_2) f_3 (\vec{x}_3) |_{\vec{x}_1 = \vec{x}_2 = \vec{x}_3 = \vec{x}} ~.
\ee
This, in turn, implies that the $\star_p$-bracket does not obey Leibnitz's rule. For any three functions of $x$, we obtain
the following result,
\ba
& & \{\{f_1 \star_p f_2 , ~ f_3 \}\}_p - f_1 \star_p \{\{f_2, ~ f_3 \}\}_p - \{\{f_1, ~ f_3 \}\}_p \star_p f_3 = \nonumber\\
& & - 6i ~ {\rm sinh} \Big[{R \over 4} (\vec{\nabla}_{x_1} \times \vec{\nabla}_{x_2})
\cdot \vec{\nabla}_{x_3} \Big] f_1 (\vec{x}_1) f_2 (\vec{x}_2) f_3 (\vec{x}_3) |_{\vec{x}_1 = \vec{x}_2 = \vec{x}_3 = \vec{x}} ~,
\ea
showing that there is an obstruction to Leibnitz rule attributed to non-associativity. Likewise, the associator does not vanish.
We find
\be
\{\{f_1 , ~ f_2 , ~ f_3 \}\}_p = - 12i ~ {\rm sinh} \Big[{R \over 4} (\vec{\nabla}_{x_1} \times \vec{\nabla}_{x_2})
\cdot \vec{\nabla}_{x_3} \Big] f_1 (\vec{x}_1) f_2 (\vec{x}_2) f_3 (\vec{x}_3) |_{\vec{x}_1 = \vec{x}_2 = \vec{x}_3 = \vec{x}} ~,
\ee
which reproduces $[x^i , ~ x^j , ~ x^k] = 3iR ~ \epsilon^{ijk}$, as expected, by specializing the result to linear
functions of $x$. We see that the obstruction to applying Leibnitz rule is one-half the associator of the three functions of $x$.
Similar formulae appear in \cite{Mylonas:2012pg} and they can be extended for more general functions in phase space.

Non-associativity leaves its mark in the description of dynamics in terms of the $\star_p$-bracket. One may still define the
analogue of a Hamiltonian vector field assigned to a function $f$, modulo
constants, as $\mathbb{X}_f = \{\{f , ~ \cdot ~ \}\}_p$, but, unlike the ordinary case, $\mathbb{X}_f$ does not act as derivation on the
$\star_p$-product algebra of functions in phase space. This has a dramatic effect on the validity of physical laws
in non-associative spaces, defying what is regarded to be common sense in normal circumstances. For example, as will be seen later,
non-associativity is responsible for the non-conservation of angular momentum and the non-closure of the algebra of rigid rotations
in $\mathbb{R}^3$ in problems with spherical symmetry.

Later, in section 5, we discuss the violation of angular
symmetry in non-associative geometry from a slightly different perspective by employing a magnetic field analogue of the commutation
relations among the coordinates and momenta\footnote{The magnetic field
analogue of non-commutativity/non-associativity, which is discussed later in detail, provides a complementary view to the problem:
it extends the well established interpretation of a constant magnetic field as non-commutative parameter to the background of
a monopole or other distributions of magnetic charge that may arise in Dirac's generalization of Maxwell theory.}.
This interchanges the role of $\vec{x}$ and $\vec{p}$, so that what is non-associative in space becomes
non-associative in momenta and vice-versa. The notion of angular momenta is inert to this interchange. Thus, angular momentum provides
a good observable to study signatures of non-associativity, either in space of in momenta, and could be used to put limits on
parameters by physical processes. A direct computation based on the $\star_p$-product is
relatively simple to perform in this case for the components of angular momentum are bilinear in $\vec{x}$ and $\vec{p}$ coordinates.
The details are left as exercise to the interested reader.
Other physical signatures of non-associativity can be found, but they will not be discussed here.

\subsection{The algebra of tachyon vertex operators}

We briefly compare the triple product \eqn{tsinghi} to the non-commutative/non-associative product among closed string tachyon
vertex operators (taken slightly off-shell) that was introduced recently in the literature, in
references \cite{Blumenhagen:2010hj} for the WZW model on $S^3$ and \cite{Blumenhagen:2011ph} for the toroidal flux model that we 
are considering here, based on conformal field perturbation theory.

These authors were led to consider the following $N$-product of periodic functions of $\vec{x}$ in a flux
background with general field strength $F^{abc}$,
\be
(f_1\, \tri_N \,  f_2 \, \tri_N \ldots \tri_N \,  f_N)(\vec{x}) = \exp\left(F^{abc} \sum_{1\le i< j < k\le N}
\!\!\!\!  \, \,\,\, \partial^{x_i}_{a}\,\partial^{x_j}_{b} \partial^{x_k}_{c} \right)
f_1(\vec{x}_1)\, f_2(\vec{x}_2) \ldots f_N(\vec{x}_N) |_{\vec{x}} ~,
\label{nproduca}
\ee
as closed string generalization of the open string non-commutative star-product. The product is defined in any number of dimensions
$n \geq 3$ using a flux form $F$. It can be specialized to three dimensions choosing $F$ to be the $R$-flux.  The restriction
appearing in the definition means that one
sets $\vec{x}_1 = \vec{x}_2 = \cdots = \vec{x}_N = \vec{x}$ after computing the action of the derivatives on the individual
functions, as in the definition of the star-product.

The $N$-product introduces an algebraic structure in the space of functions of $\vec{x}$ that differs from the
$\star_p$-product, in general. It can be easily verified that this new product gives rise to the following recursive the
relations, letting $f_N(\vec{x}) = 1$,
\be
(f_1\,\tri_N \,  f_2  \,\tri_N \,  \ldots\,  \tri_N \, f_{N-1} \, \tri_N\, 1)(\vec{x})
= (f_1\,\tri_{N-1}\,  \ldots\,  \tri_{N-1}\,  f_{N-1})(\vec{x}) ~,
\label{thenicerel}
\ee
whereas for $N=2$ it acts trivially, as it reduces to the ordinary commutative product of functions
\be
(f_1\,\tri_2 \, f_2)(\vec{x})= (f_1 \cdot f_2)(\vec{x}) ~.
\ee
None of these properties are common to the $\star_p$-product of functions that were discussed earlier.

Note, however, that the situation becomes different for $N=3$, allowing for direct comparison between $\,\tri_3$ and the
$\star_p$-product. Indeed, specializing the general definition \eqn{nproduca} to the tri-product, we obtain
\be
(f_1\,\tri_3 \, f_2\, \tri_3 \, f_3)(\vec{x}) = e^{F^{abc}\, \partial^{x_1}_{a}\,\partial^{x_2}_{b}\,\partial^{x_3}_{c}}
f_1(\vec{x}_1)\, f_2(\vec{x}_2)\, f_3(\vec{x}_3) |_{\vec{x}} ~.
\label{threebracketcon}
\ee
The result coincides with the triple $\star_p$-product of the three functions of $\vec{x}$ in the form given by
equation \eqn{tsinghi}; the other way of associating three functions of $\vec{x}$ via $\star_p$ amounts to flipping the
sign of the flux form. Thus, we observe that the structure of Lie algebra and Lie group cohomology encoded in $\star_p$,
which was discussed earlier,
in section 3, is recovered by considering the $\,\tri_3$-product of functions. It will be interesting to understand the
cohomological aspects of the $\,\tri_N$-product of functions for arbitrary values of $N$.

\subsection{Double phase space of parabolic flux model}

Although the construction we presented above is quite appealing, it is by no means complete because it misses the full phase space
structure of non-geometric string backgrounds. The main problem here is that the result is not independent from the chosen T-duality
frame. Yet, the non-geometric backgrounds are often related to geometric ones by T-duality and, thus, it is natural to expect
that the full phase structure of closed strings should include the coordinates $x^i$ as well as the dual coordinates $\tilde{x}_i$
and, similarly, the momenta $p_i$ as well as the dual momenta $\tilde{p}^i$ on equal footing.
The doubling of phase space is absolutely necessary for non-geometric backgrounds, because the monodromies mix coordinates with
dual coordinates by $O(D,D)$ transformations ($D=3$ in our case) \cite{Lust:2010iy,Andriot:2012vb}.
Thus, the formulation we have presented so far is only half of the story, as we have to deal with the full twelve-dimensional
phase space parametrized by $(x^i,\tilde x_i,p_i,\tilde p^i)$, aiming at a covariant formulation of the star-product under dualities.

The twisted Poisson structure in the double phase space of the parabolic flux model is determined by a tensor
$\Theta^{IJ}$ (here the indices take values $I,J=1,\dots ,12$) that generalizes \eqn{twistabrac} and it is given explicitly by
\begin{equation}
\Theta^{IJ}= \begin{pmatrix} R^{ijk}p_k &0& \delta^i_j&0 \\
0&0&0&\delta^j_i\\
-\delta^j_i & 0 &0&0 \\
0&-\delta^i_j & 0&0
\end{pmatrix}\, .
\label{tcichila}
\end{equation}
The blocks are labeled successively by $x, \tilde{x}, p, \tilde{p}$ and each one of them is a $3 \times 3$ matrix.
This choice reproduces the commutation relations among the coordinates, momenta and their dual
\be
[x^i , ~ x^j] = i R ~ \epsilon^{ijk} ~ p^k ~, ~~~~~~~
[x^i , ~ p^j] = i \delta^{ij} = [\tilde{x}^i , ~ \tilde{p}^j]
\label{tsiftari}
\ee
in the so-called $R$-flux frame, which looks privileged in the present formulation. Note, however, that $\Theta^{IJ}$ transforms
covariantly under $O(3,3)$ transformations. It suffices to determine the twisted Poisson structure in double phase space, using
any given frame, for all other frames simply follow by suitable duality transformations.

In practice, one can implement the dualities in double phase space by reshuffling the rows or columns of $\Theta^{IJ}$, thus
reproducing the brackets among the coordinates and momenta of the other flux backgrounds, which are summarized in Table 1 at
the end of section 2. For example, interchanging $x$ with $\tilde{x}$ and $p$ with $\tilde{p}$, we can rewrite \eqn{tcichila}
in the $H$-flux frame so that $\Theta^{IJ}$ depends explicitly on $\tilde{p}$ rather than $p$. In effect, the
cochain $\varphi_2 (g_1, g_2; \tilde{g}_1 , \tilde{g}_2)$ presented at the end of section 3.3 provides the matrix
$\Theta^{IJ}$ in double phase space in any given duality frame. More precisely, the cochains \eqn{yamawa1}, \eqn{yamawa2},
\eqn{yamawa3} and \eqn{yamawa4} are the twisted Poisson structures in the $H, ~ Q, ~R$ and $f$-flux frames, respectively,
when amended with the 2-cocycle $\vec{a}_1 \cdot \vec{b}_2 - \vec{a}_2 \cdot \vec{b}_1 +
\vec{c}_1 \cdot \vec{d}_2 - \vec{c}_2 \cdot \vec{d}_1$ of the double Heisenberg algebra ${\bf g} \oplus {\bf \tilde{g}}$
that has no effect on the associators.

With these explanations in mind, we outline the construction of the star-product in double phase space, first in the
$R$-flux frame.
We exponentiate the action of the position and momentum generators and their dual, given by \eqn{tsiftari}, and consider
the group elements
\be
U(\vec{a}, ~ \vec{b} ; ~ \vec{c}, ~ \vec{d}) = e^{i(\vec{a} \cdot \vec{x} + \vec{b} \cdot \vec{p} +
\vec{c} \cdot \vec{\tilde{x}} + \vec{d} \cdot \vec{\tilde{p}})} ~,
\ee
which satisfy the following product relation, based on Baker-Campbell-Hausdorff formula,
\ba
& & U(\vec{a}_1, ~ \vec{b}_1 ; ~ \vec{c}_1, ~ \vec{d}_1) U(\vec{a}_2, ~ \vec{b}_2 ; ~ \vec{c}_2, ~ \vec{d}_2) =
e^{-{i \over 2} (\vec{a}_1 \cdot \vec{b}_2 - \vec{a}_2 \cdot \vec{b}_1 +
\vec{c}_1 \cdot \vec{d}_2 - \vec{c}_2 \cdot \vec{d}_1)} \nonumber \\
& & ~~~~~~~~~ U \left(\vec{a}_1 + \vec{a}_2, ~ \vec{b}_1 + \vec{b}_2 - {R \over 2} (\vec{a}_1 \times \vec{a}_2);
~ \vec{c}_1 + \vec{c}_2, ~ \vec{d}_1 + \vec{d}_2 \right) .
\ea
\noindent
In turn, the associator works out to be
\ba
& & \left(U(\vec{a}_1, ~ \vec{b}_1 ; ~ \vec{c}_1, ~ \vec{d}_1) U(\vec{a}_2, ~ \vec{b}_2 ; ~ \vec{c}_2, ~ \vec{d}_2) \right)
U(\vec{a}_3, ~ \vec{b}_3 ; ~ \vec{c}_3, ~ \vec{d}_3) = e^{-i {R \over 2} ~ \vec{a}_3 \cdot (\vec{a}_1 \times
\vec{a}_2)} \nonumber \\
& & ~~~~~~~~~~~ U(\vec{a}_1, ~ \vec{b}_1 ; ~ \vec{c}_1, ~ \vec{d}_1) \left(U(\vec{a}_2, ~ \vec{b}_2 ; ~ \vec{c}_2, ~ \vec{d}_2)
U(\vec{a}_3, ~ \vec{b}_3 ; ~ \vec{c}_3, ~ \vec{d}_3) \right) .
\ea

Based on these relations, we introduce a star-product among any two functions in double phase space by combining
the $\star_p$-product \eqn{moyalia13} in $(x, p)$-space with the Moyal product \eqn{moyalia3} which is now taken in
$(\tilde{x}, \tilde{p})$-space. Skipping the intermediate steps, which just repeat themselves, we write down the final result,
\ba
& & (f_1 \star_{p, \tilde{p}} f_2) (\vec{x}, \vec{p}; \vec{\tilde{x}}, \vec{\tilde{p}}) = e^{i {R \over 2} ~ \vec{p} \cdot
(\vec{\nabla}_{x_1} \times \vec{\nabla}_{x_2})}
e^{{i \over 2} \left(\vec{\nabla}_{x_1} \cdot \vec{\nabla}_{p_2} -
\vec{\nabla}_{x_2} \cdot \vec{\nabla}_{p_1} \right)}
e^{{i \over 2} \left(\vec{\nabla}_{\tilde{x}_1} \cdot \vec{\nabla}_{\tilde{p}_2} -
\vec{\nabla}_{\tilde{x}_2} \cdot \vec{\nabla}_{\tilde{p}_1} \right)}  \nonumber\\
& & ~~~~~~~~~ \times f_1 (\vec{x}_1, \vec{p}_1; \vec{\tilde{x}}_1, \vec{\tilde{p}}_1)
f_2 (\vec{x}_2, \vec{p}_2; \vec{\tilde{x}}_2, \vec{\tilde{p}}_2) |_{\vec{x}_1 = \vec{x}_2 = \vec{x}, ~ \vec{\tilde{x}}_1 =
\vec{\tilde{x}}_2 = \vec{\tilde{x}}, ~ \vec{p}_1 = \vec{p}_2 = \vec{p}, ~ \vec{\tilde{p}}_1 = \vec{\tilde{p}}_2 = \vec{\tilde{p}}} ~.
\ea
Despite appearances, the new star-product is $O(3,3)$-invariant and, hence, independent from the
chosen T-duality frame. Here, the star-product is derived in the $R$-flux frame, thinking of $(x, p)$ as coordinates and momenta
and $(\tilde{x}, \tilde{p})$ as dual variables. Their role is completely reversed in the $H$-flux frame, in which case $p$ is
replaced by $\tilde{p}$ in the defining relations. Thus, the star-product in double phase space depends either on $p$ or
$\tilde{p}$, depending on the chosen frame, and this is indicated by $\star_{p, \tilde{p}}$ in the definition above.
Likewise, we can pass to any other frame of the toroidal flux model.

The $\star_{p, \tilde{p}}$-product introduces a non-commutative/non-associative algebraic structure on the space of functions
in double phase space. It serves as substitute to canonical quantization, treating the geometric and non-geometric phases of
the parabolic closed string flux model on equal footing. A more compact way to express the result as follows, using the
matrix \eqn{tcichila},
\begin{equation}
(f_1 \, \star_{p,\tilde{p}}\, f_2) (\vec{x}, \vec{\tilde{x}}, \vec{p}, \vec{\tilde{p}}) = e^{{i \over 2}
\Theta^{IJ}(p, \tilde{p})\,
\partial_{I} \otimes \partial_J} (f_1 \otimes f_2) |_{\vec{x}, ~ \vec{\tilde{x}}, ~ \vec{p}, ~ \vec{\tilde{p}}} ~ .
\end{equation}
As before, one can introduce a bracket $\{\{f_1 , ~ f_2 \}\}_{p, \tilde{p}} = -i(f_1 \star_{p, \tilde{p}} f_2 -
f_2 \star_{p, \tilde{p}} f_1)$ and use it to describe dynamics in the theory.

In this context, it is also interesting to examine, in general, how symmetries that may be present in $(x,p)$ space
transform away in the dual space $(\tilde{x}, \tilde{p})$, as a result of non-associativity.
The algebra of rigid rotations generated by angular momentum is a notable example that will
be discussed in the next section from a slightly different
perspective. Closely related issues are the disappearance of isometries that do not commute with
$T$-duality and the fate of supersymmetry after duality \cite{Bakas:1995hc}. We intend to return to all
these questions elsewhere.

\section{Magnetic field analogue of non-associativity}
\setcounter{equation}{0}

A twisted Poisson structure among the coordinates and momenta also appears in the formulation of point-particle
dynamics in the field of a magnetic monopole or other distributions of magnetic charge in space.
The structure of the commutators is the same, though the role of
coordinates and momenta is exchanged, compared to the non-geometric flux vacua of closed string theory. Here, we
discuss these analogies and examine the physical signatures of breaking Jacobi identity in classical as well as
quantum mechanics.
The magnetic paradigm helps us draw some lessons and find the right interpretation of non-associate structures in string theory.

\subsection{General considerations}

Let us consider the position vector $\vec{x}$ and the momentum $\vec{p} = m d\vec{x}/dt$ of a spinless point-particle with electric
charge $e$ (which can be either positive or negative) and mass $m$ moving in three dimensions under the influence of a background magnetic
field $\vec{B}$ that may vary from point to point. Then, one writes formally the following commutation relations, setting Planck's
constant equal to 1,
\be
[x^i , ~ p^j] = i \delta^{ij} ~, ~~~~~
[x^i , ~ x^j] = 0 ~, ~~~~~ [p^i , ~ p^j] = ie ~ \epsilon^{ijk} B_k (\vec{x})
\label{genamontu}
\ee
In turn, the associator among the momenta takes the form
\be
[p^i , ~~ p^j , ~~ p^k] = - e ~\epsilon^{ijk} \vec{\nabla} \cdot \vec{B} ~,
\label{dirakaki}
\ee
whereas all other combinations of associating position and momenta vanish. The associator \eqn{dirakaki} vanishes
automatically in Maxwell theory, but in Dirac's generalization of electromagnetism the result is not zero, in general, in the
presence of magnetic sources. This algebraic structure is identical to the parabolic flux model, with $\vec{x}$ and $\vec{p}$
interchanged, when $\vec{\nabla} \cdot \vec{B}$ is constant. We are also considering more general forms of magnetic field to
put our paradigm in wider context.

Note that we are not using any particular representation for the position and momenta of the point-particle
to write down the commutation relations \eqn{genamontu}. These relations are assumed to hold in all cases, including
Dirac's variant of electromagnetism based on electric-magnetics duality of the field equations with or without sources \cite{dirac}.
As such, they can be used to study the dynamics of a point-particle in a given magnetic background, in all generality.
Our discussion is based on the algebra of observables, instead of following the standard quantization rules,
exactly as it was done long time ago in
reference \cite{lipkin}. We further assume that $\vec{x}$ and $\vec{p}$ form a complete and irreducible set of observables
for the spinless point-particle moving in a static magnetic field $\vec{B} (\vec{x})$. In the applications,
$\vec{B} (\vec{x})$ will be taken to be spherically symmetric. That way, we can  exploit the role of angular momentum to
understand the apparent loss of associativity in the problem and recover the quantization of magnetic charge by
purely algebraic methods, whenever it is possible, without resorting to gauge fields with string singularities.
Dirac's quantization formula for the charges can then be regarded as necessary and sufficient condition for the realization
of all observables in the problem by self-adjoint operators acting on Hilbert space.

Within this framework, we will unveil a crucial difference between the
field of a single magnetic monopole and that of a continuous distribution of magnetic charges. Assuming that such hypothetical
objects do exist, we argue that the breakdown of associativity can be restated as failure to have a well-defined notion of angular
momentum, and conservation thereof, for a point-particle moving in the background of a general spherically symmetric field
$\vec{B} (\vec{x})$. Throughout this section we assume that the point-particle behaves as a probe that does not back-react to
the background field and that there are no magnetic currents in the problem. Thus, static magnetic fields should obey
the additional equation $\vec{\nabla} \times \vec{B} = 0$, implied by Maxwell-Dirac theory, which is certainly true for
all spherically symmetric vector fields.

\subsection{Classical motion in the field of magnetic charges}

We assume that the dynamics of the point-particle is described by $H = \vec{p} \cdot \vec{p} / 2m$, which,
furthermore, is considered to be the generator of time translations. Then, using the postulated commutation relations
\eqn{genamontu} among the
coordinates and momenta, one obtains the Lorentz force equation
\be
{d \vec{p} \over dt} = i [H , ~ \vec{p}] = {e \over 2m} (\vec{p} \times \vec{B} - \vec{B} \times \vec{p}) ~.
\label{elatora}
\ee

One may wonder about the validity of this derivation,
when associativity is at stake, because the commutator, as well as the corresponding classical bracket, will not necessarily obey
Leibnitz's rule $[A \cdot B, ~ C] = A \cdot [B, ~ C] + [A, ~ C] \cdot B$. Since $H$ involves the inner product of
two momentum vectors, any departure from Leibnitz's rule, while computing $[H, ~ \vec{p}]$, should be attributed to the associator
$[p^i , ~ p^j , ~ p^k] \sim \epsilon^{ijk}$, setting $i=j$ and summing over those repeated indices. The result is obviously zero,
justifying the derivation of our formula \eqn{elatora}.

Let us now specialize the problem and discuss the classical motion of a spinless particle of charge $e$ and mass $m$ in the
background of a radial magnetic field
\be
\vec{B} (\vec{x}) = {\vec{x} \over f(x)} ~; ~~~~~~~ x^2 = \vec{x} \cdot \vec{x} ~.
\label{tsarocha}
\ee
Spherical symmetry ensures that $\vec{\nabla} \times \vec{B} = 0$, which is consistent with the assumption
that the electrically charged point-particle is a probe that does not back-react to the magnetic field.
As such, $\vec{B} (\vec{x})$ can be regarded as static solution of Dirac's generalization of Maxwell theory with magnetic sources,
having
\be
\vec{\nabla} \cdot \vec{B} = \rho (x) ~.
\ee
The distribution of magnetic charge is continuous and spherically symmetric and its density is determined by $f(x)$ via
\be
\rho (x) = {3f(x) - x f^{\prime} (x) \over f^2} ~.
\ee
The density $\rho (x)$ is taken to be sufficiently smooth function with the exception of
a localized source of magnetic charge placed at the origin, with
$\rho (x) = 4 \pi g \delta (x)$, which corresponds to the profile function $f(x) = x^3/g$. The later configuration
is a Dirac monopole with magnetic charge $g$. In all other cases, $\rho (x)$ can be viewed as continuous
superposition of magnetic monopoles so that the total magnetic charge placed in any given region of space is obtained
by integrating $\rho (x)$.

The Lorentz force acting on the charged particle takes the same form as in Maxwell theory, according to \eqn{elatora}.
In our case, equation \eqn{elatora} specializes to
\be
{d^2 \vec{x} \over dt^2} = - {e \over m}{1 \over f(x)} \left(\vec{x} \times
{d \vec{x} \over dt} \right)
\label{ourguya}
\ee
so that the Lorentz force on the particle is proportional to its orbital angular momentum. It is possible to study the
orbit of the particle geometrically, using the Frenet-Serret relations of embedded curves in $\mathbb{R}^3$, and
find relations among the extrinsic curvature and the torsion of the curve $\vec{x}(t)$. Here, we follow a more
direct method to investigate the motion of a charged particle in the magnetic field background \eqn{tsarocha}.

The system of equations \eqn{ourguya} can be partially solved for all choices of $f(x)$. We first note that
$A = (d \vec{x} / dt) \cdot (d \vec{x} / dt)$ is a constant of motion, expressing the conservation of
energy, that follows from the orthogonality between the velocity and acceleration vectors. Also, using the orthogonality
between the position and acceleration vectors - the Lorentz force does no work - we find that $d^2 (\vec{x} \cdot \vec{x})/ dt^2 = 2A$
is also a constant expressed in terms of the energy. Integrating the last relation, we obtain the general result
\be
x^2 (t) = At^2 + D ~,
\label{perihelion}
\ee
choosing the origin of time as the moment that $x(t)$ attains its minimum value. $A$ and $D$ are positive integration constants.
It means that the particle can not come closer to the origin than a certain distance $\sqrt{D}$, which is the "perihelion" of the
orbits $\vec{x}(t)$ and is fixed by the initial conditions. This behavior is universal as it arises
for all choices of profile function $f(x)$.

A simple calculation also yields
\be
{d \over dt} \left(\vec{x} \times {d \vec{x} \over dt} \right) = \vec{x} \times {d^2 \vec{x} \over dt^2} =
- {e \over m} {1 \over f(x)} \vec{x} \times \left(\vec{x} \times {d \vec{x} \over dt} \right) =
{e \over m} {x^3 \over f(x)} {d \hat{x} \over dt} ~,
\label{intermatio}
\ee
where $\hat{x}$ denotes the unit position vector, but this equation can not be integrated for arbitrary choices of $f(x)$.
This shows how far one can go in dealing with the general case, since, to the best of
our knowledge, there are no additional integrals of motion that can be used to integrate the system \eqn{ourguya} completely.
A few more comments about the motion of the particle for generic choices of $f(x)$ will be made later.
It is an interesting problem that deserves further attention on its own.

The problem simplifies considerably when the magnetic field is sourced by a single magnetic monopole, which corresponds to
the choice $f(x) = x^3 /g$. Then, the field equations can be easily integrated, simply because there are additional constants
of motion provided by equation \eqn{intermatio}. We obtain, in particular,
\be
\vec{x} \times {d \vec{x} \over dt} - {e g \over m} \hat{x} = \vec{K} ~,
\label{moutzouria}
\ee
where $\vec{K}$ is an arbitrary constant vector fixed by the initial conditions; its components provide the three additional
integration constants for complete solvability of the equations of motion. Taking the scalar product with $\hat{x}$,
it follows that $ \vec{K} \cdot \hat{x} = -eg/m$, meaning that the angle between $\vec{K}$ and the position vector remains
constant at all times. Thus, the motion of the point-particle is entirely confined on the surface of a cone whose tip is
the location of the magnetic pole. A charged particle with energy $mA/2$  spirals on the surface of the cone staying away
from its tip by a distance $D$ or more at all time. The evolution is best described in the form
\be
{d \hat{x} \over dt} = {1 \over At^2 + D} ~ \vec{K} \times \hat{x} ~,
\ee
showing that the particle precesses around the fixed direction $\vec{K}$ with angular velocity $\vec{K}/(At^2 + D)$ that varies
with time. In this way, we reproduce a classic result first reported in reference \cite{lapidus} (but see also \cite{sivar} for a more
recent discussion of the problem and some of its generalizations).
$\vec{K}$ is nothing else but the celebrated Poincar\'e vector \cite{poincare}, which is associated to conservation of the
improved angular momentum in a Dirac monopole field, as will be seen in the next subsection.

A typical trajectory of a charged particle in a magnetic monopole field is depicted in Fig.3. The particle starts
from afar at $t = - \infty$ and begins approaching the monopole core until it reaches the "perihelion" at $t=0$ and, then,
it scatters off to infinity as $t \rightarrow + \infty$. Since the Lorentz force is normal to the surface of the cone, the
orbit is a geodesic curve on the cone. Clearly, there are no bound states in the problem. The Poincar\'e vector is pointing
up, along the axis of the cone, when $e$ is positive and down when it is negative, so that $\vec{K} \cdot \hat{x}$, which
determines the opening angle of the cone, has the appropriate sign.

\begin{figure}[h]
\centering
\epsfig{file=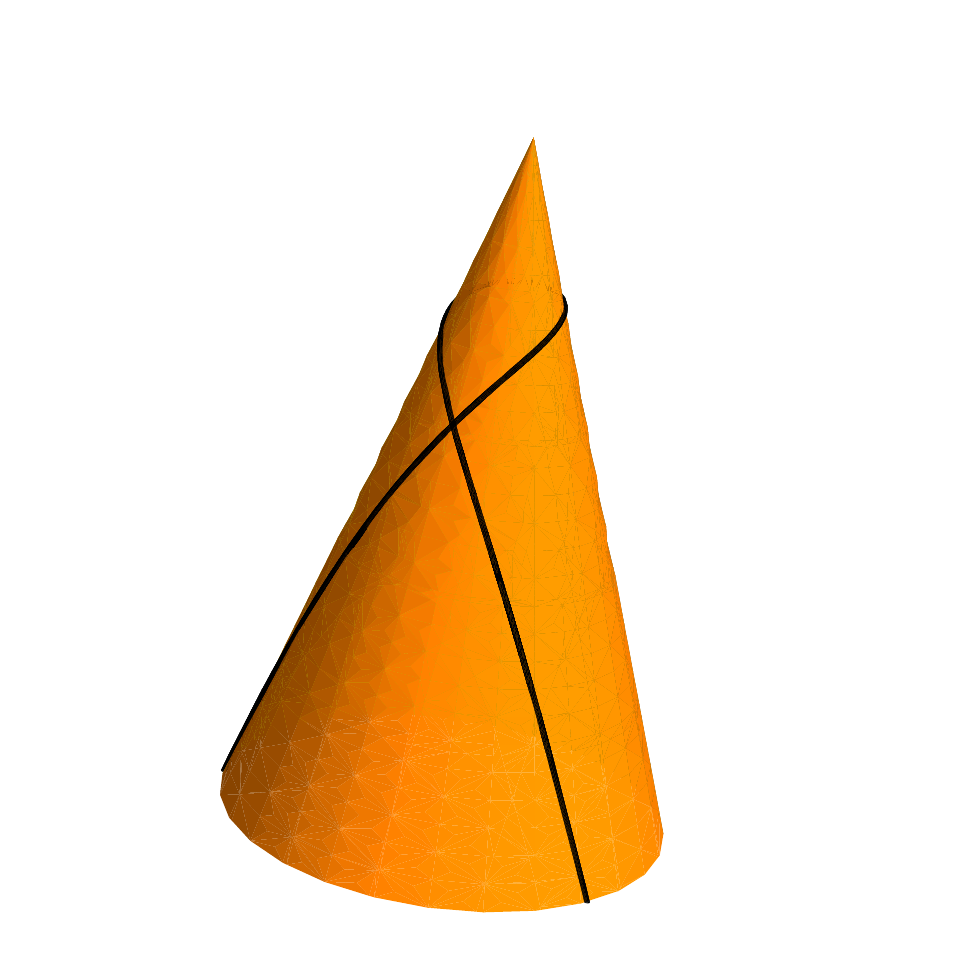,width=7cm}
\caption{\small Typical trajectory of a charged particle in the field of a magnetic monopole.}
\end{figure}

Having explained the privileged role of the magnetic monopole background, it is now possible to gain some qualitative
understanding of the more general problem \eqn{ourguya} for arbitrary choices of profile function $f(x)$. Taking advantage
of equation \eqn{perihelion}, which is valid in all cases, one can introduce suitable time reparametrization $T = T(t)$ to
transform the equations of motion \eqn{ourguya} for a general magnetic field \eqn{tsarocha} into the equations for a charged
particle in a monopole field, up to friction terms. Here, we only consider the case of a uniform constant distribution of
magnetic charge $\rho (x) = \rho$, so that $\vec{B} = \rho \vec{x} /3$, noting that all other choices of $f(x)$
can be treated similarly.

The choice $f(x) = 3/ \rho$ yields the magnetic field analogue of the $R$-flux model,
since $\vec{\nabla} \cdot \vec{B} = \rho$ is constant throughout space. Introducing $T(t)$ as
\be
{d T \over dt} = {\rho \over 3g} ~ x^3 (t) = {\rho \over 3g} ~ (At^2 + D )^{3/2} ~,
\ee
we find that the equations of motion take the following form with respect to the new time variable $T$,
\be
{d^2 \vec{x} \over dT^2} + \beta ~ {d \vec{x} \over dT} = - {eg \over m}{1 \over x^3} \left(\vec{x} \times
{d \vec{x} \over dT} \right) .
\ee
$T(t)$ is a monotonic function of $t$ extending from $-\infty$ to $+\infty$ as $-\infty < t < + \infty$ and it is
given explicitly by
\be
T (t) = {\rho \over 24 g} \Big[t(2At^2 + 5D) \sqrt{At^2 + D} + 3{D^2 \over \sqrt{A}} ~ {\rm log}
{| \sqrt{A} ~ t + \sqrt{At^2 + D} | \over \sqrt{D}} \Big]
\ee
with $T(0) = 0$. The coefficient $\beta$ of the friction term is time dependent, given explicitly by
\be
\beta (t) = {9 g \over \rho} {A t \over (At^2 + D )^{5/2}} ~,
\ee
but it can also be written in terms of the distance $x^2 = At^2 + D$ or in terms of $T$.

The motion of a charged particle in the magnetic field $\vec{B} = \rho \vec{x}/3$, which is attributed to a
uniform distribution of magnetic charge in space, is effectively described as motion in the field a single magnetic
monopole but with some friction. The friction term accounts collectively for the uniform distribution of
magnetic charge in space, treating it as viscous medium in the new formulation of the problem. In the new frame,
the acceleration of the particle is neither perpendicular to $\vec{x}$ nor to $d \vec{x} / dT$, but there is
conservation of total energy since the work done by the Lorentz force is balanced by the work done by the friction. Note, however,
that the friction term is rather peculiar in that $\beta (t)$ varies, changing sign at the "perihelion". It is positive
for $t > 0$ (i.e., $T > 0$), when the particle runs away from the effective pole, and it is negative for $t < 0$ (i.e., $T < 0$),
when the particle moves towards it. Thus, the friction behaves like an ordinary drag force, satisfying
Stoke's law, only when the particle runs away, and it vanishes at the "perihelion" as well as at spatial infinity
relative to the pole.

This peculiar behavior should be held responsible for the non-integrability of the corresponding equations of motion. The
trajectory is not confined anymore on the surface of a cone. The particle can be all over the space except from
the forbidden spherical region $x^2 < D$ surrounding the origin.

\subsection{(Non)-conservation of angular momentum}

To study the behavior of the point-particle in a spherically symmetric magnetic field $\vec{B}(\vec{x})$, in general,
we also introduce angular momentum, as
\be
\vec{J} = (\vec{x} \times \vec{p}) - \vec{C} ~,
\ee
allowing for a possible improvement term to the conventional definition of orbital angular momentum. $\vec{C}$ may
depend on both $\vec{x}$ and $\vec{p}$ and it is not determined a priori. We also make the additional assumption that
the components of position, momentum and angular momentum satisfy the following system of commutation relations
\be
[J^i , ~ x^j] = i \epsilon^{ijk} x^k ~, ~~~~~~
[J^i , ~ p^j] = i \epsilon^{ijk} p^k ~, ~~~~~~
[J^i , ~ J^j] = i \epsilon^{ijk} J^k
\label{refapointa}
\ee
from which we immediately obtain the conservation law of angular momentum, since $[H , ~ J^i] = 0$.

The philosophy of this framework, which was first advocated in reference \cite{lipkin}, is to follow as closely as possible the
conventional definitions and algebraic structures of particle dynamics, without assuming particular representations
for the observables, which may not always exist, as will be seen shortly. We only assume irreducibility of $x^i$ and $p^i$ and that 
$p^i$ act as derivations (not necessarily represented by operators acting on Hilbert space) and examine how far one can go in the
problem when there is a background field $\vec{B} (\vec{x})$
sourced by a magnetic monopole or by a spherically symmetric continuous distribution of magnetic charge.
Using the postulated commutation relations among the
angular momentum and $\vec{x}$ and $\vec{p}$, we obtain the following constraints on the improvement term $\vec{C}$,
\be
[x^i , ~ C^j] = 0 ~, ~~~~~~
[p^i , ~ C^j] = ie \left(x^i B^j - \delta^{ij} (\vec{x} \cdot \vec{B}) \right) ,
\label{improva1}
\ee
whereas the assumption that $J^i$ generate the algebra of angular momentum leads to the additional
relation\footnote{The computations are performed as if $[A \cdot B , ~ C] = A \cdot [B, ~ C] + [A, ~ C] \cdot B$.
Nowhere in these formulae there are three momenta in place of $A$, $B$ and $C$ and, thus, the corresponding associators vanish.}
\be
C^i = e x^i (\vec{x} \cdot \vec{B}) + {i \over 2} \epsilon^{ijk} ~ [C^j , ~ C^k] ~.
\label{improva2}
\ee

In Dirac's generalization of Maxwell theory, one may consider a single magnetic monopole placed at the origin of the
coordinate axes and examine the role of angular momentum to the dynamics of an electrically charged point-particle.
This provides a simple instance of the more general problem that we are posing here.
The magnetic field of a monopole is central,
\be
\vec{B} (\vec{x}) = g ~ {\vec{x} \over x^3} ~,
\ee
with $g$ being the magnetic charge of the source so that $\vec{\nabla} \cdot \vec{B} = 4 \pi g \delta (\vec{x})$.
It can be easily seen that the notion of angular momentum is well defined in the background of a monopole provided
that $\vec{J}$ includes the improvement term
\be
\vec{C} = e g ~ \hat{x}
\ee
that solves equations \eqn{improva1} and \eqn{improva2} above. This improvement term can be attributed to the angular momentum
of the electromagnetic field when the monopole core as well as the electric charge of the point-particle are regarded as fixed
sources. We recall that the combination
\be
\vec{J} = \vec{x} \times \vec{p} - eg ~ \hat{x}
\ee
first appeared in the literature at the end of 19th century (since then it became known as Poincar\'e vector \cite{poincare})
and it is the same as $m \vec{K}$ appearing in equation \eqn{moutzouria}.
It provides the conserved quantity that helps to integrate completely the
classical equations of motion of a charged point-particle in the field of a monopole, as explained earlier.

This does not yet impose any restrictions on the coefficient of the improvement term, $eg$, which remains undetermined at
the classical level. According to general theory, rigid rotations by an angle $\theta$ around any given axis $\hat{n}$ in
space are described by
\be
R (\hat{n} , \theta) = e^{-i \theta ~ \hat{n} \cdot \vec{J}} ~.
\ee
For a point-particle in a monopole field we simply have $\hat{x} \cdot \vec{J} = - eg$ and, thus, choosing the axis of
rotation as $\hat{x}$, it turns out that
\be
R (\hat{x} , \theta) = e^{-i eg ~ \theta} ~.
\ee
Single valuedness of $R (\hat{x} , \theta)$, up to a sign, requires that $eg = n \in \mathbb{Z}$ (in units of $\hbar /2$),
recovering that way Dirac's quantization condition for the electric charge $e$ relative to the strength of the magnetic
pole $g$, \cite{dirac}. This argument, which first appeared in reference \cite{wilson}, to the best of our knowledge, can be
made more rigorous, but we will not spell out the details. It shows that
the notion of angular momentum and its conservation law are well defined even in a magnetic monopole field provided
that Dirac's quantization condition is obeyed.

Let us go a step further to reach the same algebraic structure that arose in the parabolic flux models, but with $\vec{x}$ and
$\vec{p}$ interchanged. For this purpose, we choose a uniform constant density of magnetic charge in space, so that
$\vec{\nabla} \cdot \vec{B} = \rho$ is constant. It is the magnetic analogue of having uniform constant density of electric
charge in electromagnetism, which, of
course, is absurd in Maxwell's theory, but it is perfectly fine in Dirac's theory as long as single monopoles are assumed to
exist. $\vec{\nabla} \cdot \vec{B} = \rho$ is an inhomogeneous equation for the magnetic field, whose general solution is the sum of a
particular solution plus the general solution of the homogeneous equation $\vec{\nabla} \cdot \vec{B} = 0$, which can be formulated in
terms of a vector potential $\vec{A}$, as usual. A uniform distribution of magnetic charge in space yields
\be
\vec{B} (\vec{x}) = {\rho \over 3} ~ \vec{x}
\label{sperichoi}
\ee
up to solutions of the homogeneous equation, $\vec{B} = \vec{\nabla} \times \vec{A}$, which we are ignoring here\footnote{Different 
choices of $\vec{B}$ are physically distinct. The choice \eqn{sperichoi}
is the magnetic field analogue of the symmetric 2-form $B$-field \eqn{newframei}. The less symmetric solution for the
magnetic field, $\vec{B} = (0, 0, \rho x^3)$, is analogous to the $B$-field frame \eqn{lesssymme} and it leads to different
results that will not be discussed here.}.
In this case, the commutation relations \eqn{genamontu} among the position and momenta specialize to
\be
[x^i , ~ p^j] = i \delta^{ij} ~, ~~~~~
[x^i , ~ x^j] = 0 ~, ~~~~~
[p^i , ~ p^j] = ie \rho ~ \epsilon^{ijk} x^k ~,
\ee
as required for the purpose of comparison to the parabolic flux model of closed string theory. Here, non-commutativity and
non-associativity occurs in momentum space, whereas in the closed string flux models it arises among the coordinates.

It should be noted for completeness that this
particular magnetic field model was first considered in reference \cite{murat}, while searching for models of non-associative
structures based on Malcev algebras (see also \cite{murat2} for a more recent discussion of the subject unveiling links to
non-commutative and non-associative structures in string theory). Our method of investigation is mathematically different, although 
it can be regarded as complementary to theirs in some respects. Also, the motion of a point-particle in 
linear magnetic field was not considered in that work. We will not expand on the comparison here.

Placing an electrically charged particle in the background of the magnetic field \eqn{sperichoi} leads to
paradox: even though there is spherical symmetry in the problem, there is no well defined notion of angular momentum,
mind its conservation law associated to invariance under rotations in space. Simply, one cannot satisfy the assumed commutation
relations \eqn{refapointa}. Indeed, looking at the conditions
\eqn{improva1} and \eqn{improva2} imposed on the improvement term $\vec{C}$, we find that for central magnetic fields
$\vec{B} (\vec{x}) = \vec{x} / f(x)$ a solution $\vec{C} (\vec{x})$ exists if and only if $f(x) = x^3$, up to a multiplicative
constant. Thus, $\vec{J}$ is a bona fide angular momentum only in a Dirac monopole field and, apparently, there is no analogue of it
for any other choice of $f(x)$, including the magnetic field \eqn{sperichoi}. This result manifests as non-integrability
of the classical equations of motion of a charged particle in the field $\vec{B} (\vec{x}) = \vec{x} / f(x)$ for all choices
of $f(x)$ other than the profile of a monopole field, as noted earlier. After all, if angular momentum were well defined in
the magnetic field \eqn{sperichoi}, and more generally in $\vec{B} (\vec{x}) = \vec{x} / f(x)$, there will
be a quantization condition for $e$ in terms of the magnetic charge distributed in space, as in the simple case of a
single magnetic monopole. Naturally, we do not expect to have such a quantization condition for a continuous distribution of
magnetic charge in Maxwell-Dirac theory.

Summarizing the discussion so far, we note that there is an intimate relation between non-conservation of angular
momentum and non-associativity of linear momenta in the presence of the magnetic field $\vec{B} (\vec{x}) = \vec{x} / f(x)$.
The case of the Dirac monopole is very special in that the charged particle manages to escape the conflict marginally and still
allow for a well defined quantum mechanical description, as for angular momentum. More about it will be discussed later.

\subsection{Quantization in magnetic charge background}

Let us now discuss the problem from a slightly different perspective that puts the genuine
non-associative case in better context, while addressing its role in quantum mechanics. We are going to use the
cohomology of the Heisenberg algebra with cochains taking values in the space of local smooth functions of $\vec{x}$ to
revisit the general system of commutation relations \eqn{genamontu} in Maxwell as well as in Dirac's generalization of
Maxwell theory. In this context, the magnetic field is viewed as a 2-cochain that deforms the commutation relations
among the momenta, as $[p^i , ~ p^j] \sim \epsilon^{ijk} B_k (\vec{x})$.

Actually, the framework we are using here is much broader than the one discussed earlier. It can become even broader,
though it will not be necessary for the purposes of the present discussion, by considering cochains that take values in the
space of smooth functions in phase space. Specializing to
the subspace of linear functions we recover the cohomology of the Heisenberg algebra with coefficients in the algebra
itself that was extensively used earlier. We are making this generalization to accommodate all solutions
of Maxwell-Dirac theory, including the magnetic monopole and other configurations, and not just the case of linear
magnetic field.

In Maxwell theory, we solve $\vec{\nabla} \cdot \vec{B} = 0$ in terms of a vector potential
$\vec{A} (\vec{x})$, which is well defined and smooth everywhere in space, though it is not unique, as
$\vec{B} = \vec{\nabla} \times \vec{A}$. Then, we use $\vec{A} (\vec{x})$
to represent the momenta as $\vec{p} = -i \vec{\nabla} - e \vec{A}$ acting on Hilbert space. The vector potential
can be viewed as a 1-cochain in the cohomology theory of the Heisenberg with coefficients in the space of smooth
functions of $\vec{x}$. A Maxwellian magnetic field has the interpretation of 2-coboundary in the aforementioned Lie algebra
cohomology\footnote{In this context, the vector identity $\vec{\nabla} \cdot (\vec{\nabla} \times \vec{A}) = 0$ is the cocycle
condition for the 2-coboundary $\vec{B} (\vec{x})$, which is trivially satisfied, since $d^2 = 0$ in cohomology.},
described as $d\vec{A}$ in terms of the coboundary operator $d$, which, in turn, explains why the effect of
$\vec{B} (\vec{x})$ can be solely described by the minimal coupling rule with $\vec{A} (\vec{x})$.
In this case, the momenta $\vec{p}$ are well behaved self-adjoint operators everywhere and, of course, their action is
associative. A background magnetic field introduces non-commutativity among the momenta,
affecting the classical equations of motion as well as the quantum mechanical description of the point-particle.

The interpretation of the magnetic field as 2-coboundary means, mathematically, that $\vec{B} (\vec{x})$
can be moved in and out of the commutation relations by shifting the momenta generators by the
the vector potential $\vec{A} (\vec{x})$. This, however, does not trivialize its physical relevance, since, otherwise,
physics in a constant magnetic field would be the same as without it, which is, of course, incorrect. The magnetic field,
if not zero, should be present in the commutation relations among the coordinates and momenta.
What is physically irrelevant is only the freedom to perform gauge
transformations that shift the 1-cochain $\vec{A} (\vec{x})$ by a 1-coboundary term for any given choice of $\vec{B} (\vec{x})$.
The gauge transformations are derived from 0-cochains $f$, as $df$.
In more mathematical terms, it means that we are actually considering equivariant cohomology with respect to the
physically irrelevant gauge transformations.

In a Dirac monopole field the situation is different, but still it can be described in terms of the cohomology of the
Heisenberg algebra with coefficients in the space of smooth functions of $\vec{x}$. The crucial point is that there is no
globally defined vector potential for the monopole field and, hence, one proceeds in patches to avoid
string singularities of the potential. A singular choice for $\vec{A} (\vec{x})$ would take us outside the cohomology
based on smooth functions. On each patch, one introduces a vector potential that is a 1-cochain with values in the space of
smooth functions of $\vec{x}$, as before. Then, gauge transformations are used to provide the transition function on the overlap
of two patches. The patching is topologically non-trivial and it manifests in Lie algebra cohomology as having a magnetic
deformation of the Heisenberg algebra driven by a 2-cochain that is not a coboundary anymore. Acting on it with the coboundary
operator does not give zero, but $\vec{\nabla} \cdot \vec{B}$ instead, which appears to obstruct the Jacobi identity
among the momenta. The obstruction can be viewed as a 3-coboundary that is localized at a point, where the magnetic monopole sits.

Representing the momenta as $\vec{p} = -i \vec{\nabla} - e \vec{A}$ is correct only in patches and it can not be
extended everywhere in space with the same vector potential without hitting string singularities. These singularities,
which emanate from the monopole core take us outside the space of local functions and they are physically unacceptable.
Thus, the Jacobi identity holds only in patches. One may still try to think of the momenta as linear operators acting on the
Hilbert space of functions on $\mathbb{R}^3$, but their domain should be restricted on the given patch, where the representation
$\vec{p} = -i \vec{\nabla} - e \vec{A}$ is valid. Trying to extend their validity over the entire space with the use of
a singular vector potential creates a problem with the self-adjointness of the operators that are supposed to represent the
momenta. To remedy the situation, one tries to choose some self-adjoint extension by imposing appropriate boundary conditions
on the allowed wave functions. Since the momenta act by shifting the argument of the wave-functions, they take us outside the
domain that is available for them on any given patch. Requiring that the wave-functions vanish at the location of the magnetic
monopole resolves the problem and provide us with the necessary self-adjoint extension for the momenta operators.

Thus, the apparent violation of Jacobi identity washes away when the associator among the momenta is viewed
as an operator equation on the Hilbert space of functions that vanish at the origin. This
boundary condition is also consistent with the classical behavior of a charged point-particle in a monopole field, which moves
in space by avoiding the vicinity of the pole, as noted earlier. The resolution of the problem, as it stands,
does not rely on any quantization condition. Recall, however, that
the quantum mechanical description of a point-particle in the field of a magnetic monopole is consistent provided that the
physical wave functions are single-valued. This additional requirement leads to Dirac's quantization condition
$e g = n \in \mathbb{Z}$ (in units of $\hbar /2$) by a standard argument based on patches \cite{dirac}. The methodology
and results are analogous to the previous discussion: the existence of the Poincar\'e vector and the conservation of angular
momentum did not rely on any quantization of charges; it was the additional requirement that rigid rotations should be
represented by single-valued transformations that led to the Dirac's quantization condition by operator methods.

Other descriptions of the same problem that employ self-adjoint extensions of operators and the role of rotations in Hilbert
space have also appeared in the literature (see, for instance, the old paper \cite{hurst}). In more recent years, the
problem was revisited using the Abelian group of translations of $\mathbb{R}^3$ and it was found that the Dirac monopole
leads to a real-valued non-trivial 3-cocycle that obstructs associativity of the product law of finite translations
\cite{jackiw, berny, allrest} (but see also \cite{carey} for some important followup work).
This is a magnetic version of our earlier description of parabolic flux models in terms of real-valued cohomology of the
algebra of translations, as alternative to using cohomology of the Heisenberg algebra with cochains having non-trivial
coefficients (i.e., taking values in the Heisenberg algebra itself or in the more general space of smooth functions of $\vec{x}$
that was considered above). Despite appearances, associativity is regained in the background of a monopole field after
imposing Dirac's quantization condition. The 3-cocycle becomes an integer multiple of $2 \pi$ and its effect disappears
from the associator of any three group elements.

The situation changes drastically in the magnetic field background $\vec{B} (\vec{x}) = \rho \vec{x} / 3$ assigned to
a continuous uniform distribution of magnetic charge in space. Definitely, we do not expect to be able to use patches to
describe the effect of this magnetic field by gauge connections, since there are going to be string singularities all over
space. Also, we do not
expect to have any quantization condition for the continuous distribution of magnetic charge. Thus, the problem is
genuine non-associative and any one of the previous methods and techniques does not help to deal with it in quantum mechanics.
If we were applying the same line of thought, as before, the wave-functions would be zero everywhere in order to maintain
self-adjointness of the momenta operators. The breakdown of associativity is in general inconsistent with the use of linear
operators acting on a common dense invariant domain of a Hilbert space, showing that there is no way to treat the problem at
hand by conventional means. The breakdown of angular symmetry that was discussed earlier shows yet another face of the problem
both in classical and quantum mechanics.

An interesting point of view regarding the interpretation of non-associativity among the momenta was advocated by Carey
et al \cite{carey} (see, in particular, the third paper in that series of references). It applies to the magnetic field
$\vec{B} (\vec{x}) \sim \vec{x}$ that we are considering here, but it does not seem to extend to more general cases
when the 3-cocycle is not anymore constant.
In that approach, motivated by the relations $[x^i, ~ p^j] = i \delta^{ij}$, one thinks of the momenta as
derivations defined by commutators on the algebra of observables. The problem arises when one wants to represent the
momenta by operators on a Hilbert space rather than derivations on an algebra. The 3-cocycle obstructs precisely
this correspondence and there is nothing one can do about it. Similar considerations apply to the parabolic flux
model by exchanging the role of position and momentum.

The star-product among the phase space functions of a point-particle moving in magnetic field should substitute
for canonical quantization. It is provided by the magnetic field analogue of the $\star_p$-product introduced earlier.
Classical mechanics should then be described by the twisted Poisson bracket $\{f_1 , ~ f_2\}_p$ that arises by taking
the limit $\hbar = 0$ of the $\star_p$-bracket $\{\{f_1 , ~ f_2\}\}_p$ introduced earlier in section 4.2, replacing
$p$ by $x$ in the magnetic field models.

Other examples with non-constant $\vec{\nabla} \cdot \vec{B}$, and likewise other examples of non-geometric closed string
backgrounds with non-constant fluxes, are interesting to consider, but they will not be discussed here.

\section{Conclusions and discussion}
\setcounter{equation}{0}

We presented a systematic description of the parabolic flux model of closed string theory using the deformation
theory of Lie algebras and their cohomology. A non-associative star-product on the space of phase space functions
was constructed from first principles, complementing some earlier work on the subject \cite{Mylonas:2012pg}. It
was further extended to the double field theory phase space of these models encompassing all coordinates, momenta, and
their dual on equal footing. That way, we obtained a unified description of all faces of the parabolic flux model, which
arise by a sequence of T-duality transformations from toroidal compactifications of string theory with constant 3-form flux.
These models expose very nicely and in simple terms the interplay between coordinates and momenta and their dual
counterparts in the geometric and non-geometric faces of string vacua.

We also presented and expanded on a rather
peculiar problem regarding the motion of a charged particle in the field of magnetic charges and compared it to the
parabolic flux model of string theory in the canonical formalism. Such toy models help to understand better the
special features of non-commutativity/non-associativity in the classical and quantum domain and bring to light
the role of non-associativity to symmetries and conservation laws that would have existed in normal circumstances.
The non-associative $\star_p$-product, and its double phase space generalization, can be used as substitute for
canonical quantization of the parabolic flux model and its magnetic field analogue (after introducing the necessary
interchange of position and momentum). It remains to be seen how precisely the approach we followed here can be
fully accommodated and expanded in the general framework of double field theory that is being developed as new
tool for reformulating string theory.
Connections of non-geometric flux models to (non)-associativity and magnetic monopoles also appear in the context
of $D$-brane and matrix models of open string theory (see, for instance, \cite{corna} and \cite{athanasios}).

Interpreting the obstruction to Jacobi identity as coboundary in Lie algebra cohomology with non-trivial coefficients,
suggests that the appropriate mathematical structure underlying the parabolic flux models (and possibly other non-geometric
models of closed string theory) may be that of a strongly homotopy Lie
algebra. We refer the interested reader to \cite{jimbo}, and references therein, for a physics oriented review of the subject,
but see also reference \cite{markl} for explaining the role of homotopy algebras to deformation quantization based on the original
ideas of Kontsevich, \cite{kontsev}, which were also used by the authors of reference \cite{Mylonas:2012pg} to construct the
$\star_p$-product of functions on phase space. According to them, but it also follows implicitly from our work, the
appropriate underlying structure is that of a 2-algebra, whereas for other non-geometric string backgrounds
generalizations to higher algebras might be needed. Further exploration of these possibilities are left open to future work.

In an interesting series of papers \cite{bouwknegt1, bouwknegt2, bouwknegt3, bouwknegt4, mathai},
Bouwknegt and collaborators have also considered the case of dualizing along the
cycles of tori carrying a 3-form flux. Their approach is applicable to principal torus bundles $M \times T^n$ over a base
manifold $M$, which can be general, and they employ techniques mainly from the cohomology theory of differential forms
to describe in purely algebraic terms the geometric as well as the non-geometric faces of flux tori. They also used
the general mathematical machinery of operator algebras, generalizing the notion of $C^{\star}$-algebras, to provide
non-commutative and/or non-associative products of functions for the non-geometric faces of the theory (it is what they
call non-commutative and/or non-associative tori). Our approach, compared to theirs, is more elementary and (in our view)
more illuminating, as it starts from first principles employing the basic commutation relations among the coordinates and
momenta that were extracted by conformal field theory methods. Further comparison of the two approaches might
prove useful and sharpen our current understanding of the subject.

Other classes of non-geometric string models are also interesting to consider along the lines of the present work. They
include the flux models discussed in references \cite{Lust:2010iy, Condeescu:2012sp}, which have elliptic monodromies and the
gluing conditions of the fibre, when going around the base, are given by finite order $O(D,D)$ transformations. They are
technically much more involved in that their canonical formulation requires making use of the full phase space of
coordinates, momenta, and their dual. Their commutation relations resemble those of the parabolic model, but the
deformation terms that lead to non-associativity are non-linear functions of the coordinates and momenta. One can still
use cohomological methods to write down and characterize the associator of the elliptic models, but this time the
cochains should take values in a larger module of the double Heisenberg algebra, namely in its universal enveloping algebra
or the algebra of smooth functions on double phase space.
Repeating the construction of the non-associative star-product is not an easy task because one needs the full
non-associative analogue of Baker-Campbell-Hausdorff formula to perform the calculations. It should also be noted
that there is no obvious magnetic field analogue of the elliptic as well as the more general non-geometric backgrounds
that are currently available in the literature, because one cannot simply model them with half of double field theory
phase space. These remarks indicate some of the differences with the parabolic flux model, which is the simplest of all
non-geometric backgrounds, and they can be substantiated, hoping that a more universal picture emerges from their study.
Work along these lines is currently in progress.

\vskip1cm

\section*{Acknowledgements}
This work was partially supported by the ERC Advanced Grant "Strings and Gravity"
(Grant.No. 32004) and by the DFG cluster of excellence "Origin and Structure of the Universe".
This research is also implemented (I.B.) under the "ARISTEIA" action of the
operational programme education and lifelong learning and is co-funded
by the European Social Fund (ESF) and National Resources.
The first author (I.B.) is grateful to the hospitality extended to him at the Arnold
Sommerfeld Center for Theoretical Physics in Munich during the course
of this work as well as to the hospitality of the theory group at CERN. We thank
Ralph Blumenhagen, Andreas Deser, Ioannis Florakis, Falk Hassler and Olaf Hohm for fruitful discussions
and collaboration on related matters. Special thanks are also due to Jim Stasheff
for some useful correspondence on the mathematical aspects of non-associative structures
and Christos Sourdis for assistance in preparing the figures.


\appendix
\section{Cohomology of Lie algebras and Lie groups}
\setcounter{equation}{0}

In this appendix we present the basic elements of cohomology theory for Lie algebras and Lie
groups that are needed in the main text. A good reference is the original work of Chevalley and
Eilenberg \cite{eilenberg} (but see also \cite{knapp, kebro}), where further details on the subject can be found.

{\bf I. Lie algebra cohomology}: The main objects in this theory are the so called $r$-cochains, denoted by
$C^r ({\bf g} , ~ V)$, which are totally anti-symmetric multi-linear maps $c$ from a Lie algebra
${\bf g}$ with values in a prescribed module $V$ of ${\bf g}$.
Thus, $V$ is a vector space carrying a representation $\pi$ of the algebra so that
$\pi (X) V \subseteq V$ and $[\pi (X) , ~ \pi (Y)] V = \pi ([X, Y]) V$ for all elements $X, Y \in {\bf g}$.
An $r$-cochain corresponds to an element of the product space $V \otimes X_{i_1}^{\star} \wedge X_{i_r}^{\star} \wedge
\cdots \wedge X_{i_r}^{\star}$, where $X_i^{\star}$ are linear maps from the algebra ${\bf g}$ to the reals,
i.e., $X_i^{\star}$ are elements of the dual algebra ${\bf g}^{\star}$ and, therefore, according to definition,
$c (X_{i_1} , X_{i_2}, \cdots , X_{i_r}) \in V$ for all $X_i \in {\bf g}$. We set $C^0 ({\bf g} , ~ V) = V$.

If $V = \mathbb{R}$, so that $\pi$ is the trivial representation of ${\bf g}$, the $r$-cochains are said to have
trivial coefficients and we will speak of the algebra cohomology with trivial (i.e., real-valued) coefficients.
If $V$ coincides with the algebra itself, i.e., $V = {\bf g}$, the representation $\pi$ is the adjoint
representation, i.e., $\pi (X) Y = {\rm Ad}_{X} Y = [X, ~ Y]$ and we will speak of cochains and cohomology
with values in the algebra itself. Other choices of $V$ are also commonly
used in the literature depending on circumstances and the applications one is considering.

The Lie algebra cohomology in defined in terms of the coboundary operator $d$, which is a linear map
$d: ~ C^r ({\bf g}, V) \rightarrow C^{r+1} ({\bf g}, V)$ acting in all generality on any $c \in C^r ({\bf g}, V)$,
as follows:
\ba
dc(X_{i_1}, X_{i_2}, \cdots , X_{i_{r+1}}) & = & \sum_{1 \leq k < l \leq r+1} (-1)^{k+l}
c([X_{i_k} , X_{i_l}], X_{i_1}, \cdots , \hat{X}_{i_k} , \cdots , \hat{X}_{i_l} , \cdots , X_{i_{r+1}}) \nonumber\\
& & + \sum_{k=1}^{r+1} (-1)^{k+1} \pi (X_{i_k}) c (X_{i_1}, X_{i_2}, \cdots , \hat{X}_{i_k}, \cdots ,
X_{i_{r+1}}) ~.
\ea
Hatted $X_i$'s denote omitted elements from the corresponding entries of the $r$-cochain $c$ used in the
definition. If the cochains have trivial coefficients, all terms in the second line will be
obviously zero.

The coboundary operator squares to zero, i.e., $d^2 = 0$, as can be checked directly by acting on any
$c \in C^r ({\bf g}, V)$. As such, $d$ defines a graded differential complex
\be
\cdots \stackrel{d}{\longrightarrow} C^{r-1} ({\bf g}, V) \stackrel{d}{\longrightarrow} C^r ({\bf g}, V)
\stackrel{d}{\longrightarrow} C^{r+1} ({\bf g}, V) \stackrel{d}{\longrightarrow} \cdots
\ee
called the Chevalley-Eilenberg complex of the algebra ${\bf g}$ with values in $V$. A $r$-cochain $c \in C^r ({\bf g}, V)$ is
called closed (or cocycle) if $dc = 0$ and it is called exact (or coboundary) if there is an element $\theta \in C^{r-1} ({\bf g}, V)$
such that $c= d\theta$. The Lie algebra cohomology with values in $V$ is defined by taking the quotient of the kernel with
the image of $d$, for all $r$, as follows:
\be
H^r ({\bf g}, V) = {{\rm ker} ~ d: ~ C^r ({\bf g}, V) \rightarrow C^{r+1} ({\bf g}, V) \over {\rm im} ~ d: ~
C^{r-1} ({\bf g}, V) \rightarrow C^{r} ({\bf g}, V)} ~.
\ee
If $H^r ({\bf g}, V)$ is a non-trivial group, there will be $r$-cocycles that are not coboundaries; otherwise, if
the $r$-th cohomology is trivial, all $r$-cocycles will be coboundaries.

Of particular interest is the second cohomology group $H^2 ({\bf g}, \mathbb{R})$ that is isomorphic to the space
of equivalence classes of central extensions of ${\bf g}$. They are consistent with the Jacobi identity, since the
2-cocycle condition for $V= \mathbb{R}$ reads
\be
c([X_{i_1} , X_{i_2}], ~ X_{i_3}) + c([X_{i_2} , X_{i_3}], ~ X_{i_1}) + c([X_{i_3} , X_{i_1}], ~ X_{i_2}) = 0 ~.
\ee
In this case, the 2-cocycle $c (X_{i_i}, X_{i_2})$ will be exact if there is a 1-cochain $\theta (X)$ such that
$c (X_{i_i}, X_{i_2}) = \theta ([X_{i_i}, ~ X_{i_2}])$.
Likewise, the third cohomology group with trivial coefficients accounts for possible violations of the Jacobi
identity by (what are often called in the terminology of quantum field theory) $c$-number terms, since the
3-cocycle condition for $V = \mathbb{R}$ reads
\ba
& & c([X_{i_1} , X_{i_2}], ~ X_{i_3}, ~ X_{i_4}) - c([X_{i_1} , X_{i_3}], ~ X_{i_2}, ~ X_{i_4}) +
c([X_{i_1} , X_{i_4}], ~ X_{i_2}, ~ X_{i_3}) + \nonumber\\
& & c([X_{i_2} , X_{i_3}], ~ X_{i_1}, ~ X_{i_4}) - c([X_{i_2} , X_{i_4}], ~ X_{i_1}, ~ X_{i_3}) +
c([X_{i_3} , X_{i_4}], ~ X_{i_1}, ~ X_{i_2}) = 0
\ea
whereas if $c(X_{i_1} , X_{i_2},  X_{i_3}) = \theta ([X_{i_1} , X_{i_2}], ~ X_{i_3}) +
\theta ([X_{i_2} , X_{i_3}], ~ X_{i_1}) + \theta ([X_{i_3} , X_{i_1}], ~ X_{i_2})$ for an appropriately chosen
2-cochain $\theta (X_{i_1}, X_{i_2})$, then, the 3-cocycle will be exact.

In Chevalley-Eilenberg cohomology with $V = {\bf g}$, the second cohomology group $H^2 ({\bf g}, {\bf g})$
is identified with the space of non-trivial infinitesimal deformations of ${\bf g}$ by the algebra itself, whereas
the third cohomology group $H^3 ({\bf g}, {\bf g})$ describes the obstructions to integrating such infinitesimal
deformations.

{\bf II. Lie group cohomology}: For any given Lie group $G$ we consider the Abelian group of all functions from $G^r$
($r$ copies of $G$) to
a $G$-module $V$. These elements are called {\em inhomogeneous} $r$-cochains of the group $G$ with coefficients in
$V$ and they are denoted $\varphi \in C^r (G, V)$. Thus, one has $\varphi (g_1 , g_2 , \cdots , g_r) \in V$ for all
elements $g_1 , g_2 , \cdots , g_r \in G$.

The coboundary operator $d : C^r (G, V) \rightarrow C^{r+1} (G, V)$ acting on inhomogeneous cochains is defined
for all $r$ as follows,
\ba
d \varphi (g_1 , g_2 , \cdots , g_{r+1}) & = & \pi (g_1) \varphi (g_2 , g_3 , \cdots . g_{r+1}) + \nonumber\\
& & \sum_{k=1}^r (-1)^k \varphi (g_1 , \cdots , g_{k-1} , g_k g_{k+1} , g_{k+2} , \cdots , g_{r+1}) + \nonumber\\
& & (-1)^{r+1} \varphi (g_1 , g_2 , \cdots . g_r) ~,
\ea
where $\pi (g)$ is a given representation of $G$ on $V$. The coboundary operator is nilpotent, i.e., $d^2 = 0$,
and as such it defines a complex
\be
\cdots \stackrel{d}{\longrightarrow} C^{r-1} (G, V) \stackrel{d}{\longrightarrow} C^r (G, V)
\stackrel{d}{\longrightarrow} C^{r+1} (G, V) \stackrel{d}{\longrightarrow} \cdots
\ee
as in the case of Lie algebras. In analogy with that case, we define the cohomology groups $H^r (G, V)$ as the
quotient
\be
H^r (G, V) = {{\rm ker} ~ d: ~ C^r (G, V) \rightarrow C^{r+1} (G, V) \over {\rm im} ~ d: ~
C^{r-1} (G, V) \rightarrow C^{r} (G, V)} ~,
\ee
calling a $\varphi \in C^r (G, V)$ closed (or $r$-cocycle) if $d \varphi = 0$ and exact (or coboundary) if there is
an $(r-1)$-cochain $\phi$ such that $\varphi = d \phi$.

Thus, for example, in Lie group cohomology with trivial coefficients, $V = \mathbb{R}$, the 2-cocycle condition
$d \varphi (g_1, g_2, g_3) = 0$ reads
\be
\varphi (g_2, g_3) - \varphi (g_1 g_2, g_3) + \varphi (g_1 , g_2 g_3) - \varphi (g_1 , g_2) = 0 ~.
\ee
The 2-cocycle is exact if there is a real-valued 1-cochain $\phi$ so that $\varphi (g_1 , g_2) = d \phi (g_1 , g_2) \\
= \phi (g_2) - \phi (g_1 g_2) + \phi (g_2)$. For $V = \mathbb{R}$, the second cohomology group $H^2 (G, \mathbb{R})$
is in one-to-one correspondence with the central extensions of $G$ by $\mathbb{R}$. More generally, if $V$ carries
a non-trivial representation of $G$, $H^2 (G, V)$ classifies the extensions of $G$ by the module $V$ in which the
induced action of $G$ on $V$ by inner automorphisms agrees with the given action.

Likewise, the 3-cocycle condition $d \varphi (g_1 , g_2 , g_3 , g_4) = 0$ in Lie group cohomology with trivial coefficients
assumes the form
\be
\varphi (g_2 , g_3 , g_4) - \varphi (g_1 g_2, g_3 , g_4) +
\varphi (g_1 , g_2 g_3 , g_4)
- \varphi (g_1 , g_2 , g_3 g_4) + \varphi (g_1 , g_2 , g_3) = 0 ~.
\ee
The 3-cocycle is exact if there is a real-valued 2-cochain $\phi$ so that $\varphi (g_1 , g_2 , g_3) = d \phi (g_1 , g_2 , g_3) =
\phi (g_2, g_3) - \phi (g_1 g_2 , g_3) + \phi (g_1 , g_2 g_3) - \phi (g_1 , g_2)$.

We are making use of all these conditions in the main text, adopting the definition of Lie group cohomology based on
inhomogeneous cochains, as above.

Equally well, we could have adopted an alternative definition of group cohomology that is often used
in the literature, based on {\em homogeneous} cochains, which we denote by $\omega$ to distinguish them from the inhomogeneous
cochains $\varphi$ used earlier,
\be
\pi (g) \omega (g_1, g_2 , \cdots , g_r) = \omega (gg_1 , gg_2, \cdots , gg_r) ~.
\ee
Then, there is a coboundary operator $\delta$ acting on homogeneous cochains as
\be
\delta \omega (g_1, g_2, \cdots , g_{r+1}) = \sum_{i=1}^{r+1} (-1)^k  \omega (g_1, g_2, \cdots , \hat{g}_k,
\cdots , g_{r+1})
\ee
and it squares to zero, $\delta^2 = 0$. The relation to group cohomology, as defined before, is established through
the identification $\varphi_1 (g) = \omega_2 (1, g)$, $\varphi_2 (g_1, g_2) = \omega_3 (1, g_1, g_1 g_2)$,
$\varphi_3 (g_1 , g_2 , g_3) = \omega_4 (1, g_1 , g_1 g_2, g_1 g_2 g_3)$, etc, noting that
$d \varphi_1 (g_1 , g_2) = \delta \omega_2 (1, g_1 , g_1 g_2)$, $d \varphi_2 (g_1, g_2 , g_3) =
\delta \omega_3 (1, g_1, g_1 g_2, g_1 g_2 g_3)$ and so on. There is a shift by 1 in the index characterizing the
cochains, which has to be accounted in order to establish the equivalence of the two frameworks, even for cochains with trivial
coefficients.

\end{document}